\documentclass[a4paper,11pt]{article} 
\usepackage{rotating}
\usepackage{lscape}
\usepackage{amsfonts}
\usepackage{amssymb}
\usepackage[centertags]{amsmath}
\usepackage{graphicx}
\usepackage{geometry}
\providecommand{\U}[1]{\protect\rule{.1in}{.1in}}

\usepackage{amsmath,mathtools,amssymb,natbib,bm,array,tabularx,booktabs,caption,fixltx2e,multirow, longtable,rotating,adjustbox,hhline,multirow}
\usepackage[colorlinks=true,linkcolor=black, citecolor=blue, urlcolor=blue]{hyperref}
\usepackage{lineno}
\usepackage[toc]{appendix}

\begin{document}


\title{The determinants of bank loan recovery rates in good times and bad - new evidence}

\author{Hong Wang\thanks{Department of Econometrics and Business Statistics, Monash University, Australia, email:\textit{hong.s.wang@monash.edu}.}, Catherine S. Forbes\thanks{Department of Econometrics and Business Statistics, Monash University, Australia, email:\textit{catherine.forbes@monash.edu}.}, Jean-Pierre Fenech\thanks{Department of Banking and Finance, Monash University, Australia, email:\textit{jeanpierre.fenech@monash.edu}.} \ and John Vaz\thanks{Department of Banking and Finance, Monash University, Australia, email:\textit{john.vaz@monash.edu}.},}
\maketitle

\begin{abstract}
	We find that factors explaining bank loan recovery rates vary depending on the state of the economic cycle. Our modeling approach incorporates a two-state Markov switching mechanism as a proxy for the latent credit cycle, helping to explain differences in observed recovery rates over time. We are able to demonstrate how the probability of default and certain loan-specific and other variables hold different explanatory power with respect to recovery rates over `good' and `bad' times in the credit cycle. That is, the relationship between recovery rates and certain loan characteristics, firm characteristics and the probability of default differs depending on underlying credit market conditions. This holds important implications for modelling capital retention, particularly in terms of countercyclicality.\\

\emph{\bigskip}


\emph{\bigskip}


\end{abstract}


\section{Introduction} \label{section_intro}
Loan defaults are inevitable events within a bank's loan book. Credit risk management processes require banks to accurately model loan default probabilities and subsequent recovery rates (RRs, hereafter). These models are a key compliance requirement for banks subscribing to the Advanced Internal Ratings Based (AIRB) models. Furthermore, the latest International Financial Reporting Standards on Financial Instruments (IFRS 9) in particular requires entities to reflect on a default probability based on best available forward looking information. An accurate understanding of RR performance over time is critical for banks, and could potentially result in more efficient use of capital. In this paper, we study the interaction of borrower characteristics, loan features and macroeconomic conditions together with other key criteria with respect to probability of default (PD, hereafter) and RRs across credit cycles. 

Several previous studies have investigated the determinants of RRs. See, for example, \cite{altman2005link}, \cite{acharya2007does}, \cite{bruche2010recovery}, \cite{khieu2012determinants} and \cite{altman2014ultimate}. However, the systemically time-varying PD and RR reaction to different credit and economic cycles has not been captured. Most studies assume a constant association between PD and RR, potentially leading to an inaccurate assessment of RR risk (\cite{resti2002new} and \cite{altman2005link}).

In addition, most of the contemporary literature concerning RR determinants does not focus specifically on bank loans. \cite{altman2014ultimate}, using the same set of determinants to study bank loans and corporate bonds, combine bank loans with corporate bonds, whereas \cite{mora2015creditor} investigates corporate bonds only. Bank loans are fundamentally different to other securities; typically, they are senior to traded corporate debt. Due to the different repayment hierarchy, this tends to make bank loan RRs higher. Furthermore, a bank generally has much greater access to customers’ financial information than other types of investor, forcing covenant compliance if any financial ratios or loan covenants are breached. Therefore, given their access to non-public information, banks are more likely to enforce bankruptcy than other key stakeholders, and hold more power over borrower firms with respect to RRs. Additionally, banks may gain access to underlying assets as their fixed/floating charges allow them to be paid before other creditors.

\cite{papke1993econometric} and \cite{khieu2012determinants} studied bank loans, but do not examine RRs through the 2007/08 financial crisis and beyond. They also impose parametric assumptions/constraints which are embedded within the models they employ. The limitation of such an approach is that it assumes distributions for the data that may be quite different from observed RRs. Furthermore, a quasi-likelihood method is employed. The RRs are modelled using a Bernoulli likelihood that does not naturally accommodate observed RRs that fit inside the unit interval. This approach is replicated by \cite{khieu2012determinants}, who also employ a linear modelling approach where the errors are effectively assumed to be normal - an assumption contrary to the observed RR distributions.

Most existing studies employ only static models, hence they do not account for systemically time-varying changes in the RR, arising from the economic environment. Furthermore, as discussed by \cite{altman2005link}, the negative relation between PD and RR needs to be assessed under `good times' and `bad times' as so far it is not clear how such interactions change under alternative economic conditions/cycles. The IFRS 9 implementation scheduled for 2018 creates an increased imperative to investigate this relationship.

In view of the above, this paper develops a dynamic predictive model for bank loan RRs, allowing for good and bad times. This enables us to ascertain whether variable relations are consistent over time, while accounting for the distinctive features of the empirical distribution of the data. We also account for a range of other relevant factors in a dynamic framework, as such variables have only previously been considered as RR predictors in static contexts. Here, however, the predictors are conditional upon the underlying economic and credit cycle, which in turn we characterize in line with \cite{bruche2010recovery} with two distinct states - good and bad.

Utilizing data from Moody's Ultimate Recovery Database, this study focuses on defaulted bank loans between 1987 and 2015 in the United States (US). In order to manage the latent economic states and the flexible nature of the empirical RR distribution, a Bayesian inferential methodology is developed, exploiting the hierarchical structure, along the lines of \cite{kim1999state}. Moreover, due to concerns regarding the large number of available predictors, and the fact that many of these regressors may be correlated, we incorporate a least absolute shrinkage and selection operator (LASSO) prior for the regression components. The inferential results from the dynamic model are subsequently compared to the available static versions, with new insights reported, contributing further to the literature.

Overall, we find more significant loan characteristics during good times. Conversely, during bad times, only certain collateral determinants are related to RRs. This finding reinforces the notion that not all of a firm’s assets facilitate a full loan recovery, with inventory and accounts payable more likely to achieve such an objective. This has consequences for discounts that banks apply to assets offered as collateral. The size of the discount should, we argue, not only depend on the riskiness and liquidity of the assets being offered, but also on the credit cycle.

The type of recovery process is not directly related to RRs in either cycle, however, when the time to emerge is considered alongside prepackaged recovery processes, some interesting results are found. Banks are found to have lower RRs during a bad cycle when there is no prepackaging. Conversely, recoveries in a good cycle, result in a higher RR, suggesting banks need to be mindful of the likely resolution time involved during such processes at the origination time.

We find RRs are significantly affected by the condition of the economy's countercyclicality and the borrowers’ characteristics. \cite{khieu2012determinants} report no association; however, as we control for cycles, we find larger firm size is associated with reduced RRs during bad times and the opposite during good times. Our finding highlights the importance of carefully considering the definition of idiosyncratic risk, particularly the tail risk of the bank loan loss distribution. During bad times, banks need to give greater weight to their tail-risk forecasts, covering any unexpected losses from their large customers by allocating further economic capital. Once again, such findings have an impact on IFRS 9 implementation with respect to forward-looking judgments.

Finally, in line with \cite{khieu2012determinants}, we use the all-in-spread (AIS) measure as our proxy for PD and report a statistically negative association only during bad cycles. PD and RR have historically been found to be negatively related. Although several important advances have previously been made using credit risk modelling techniques, at the individual loan level the relation between PD and RR has remained an open question. Our study provides evidence of a negative relationship during bad times only, suggesting the presence of systemic time variation in this context.

Overall, such findings have a direct impact on a bank's loan loss distribution, which is a vital component for determining capital allocation. The default loan recovery process suggests loan, recovery, borrower, economic and PD features need to be dynamically managed for banks to optimally allocate capital across credit cycles. Clearly then, debt recovery is time-varying and the potential risk of not accurately addressing such variables in the credit risk process will lead to potentially under- or over-providing for future loan losses.

These findings support the Basel III framework's recommendations for the use of countercyclical buffers, creating an environment where the banking sector is protected from periods of excessive aggregate credit growth. So capital buffers assist against such a build-up phase and help the banks’ going concern when RRs underperform. Conversely, during bad times, capital buffers are essential, as the supply of credit may be curtailed by regulatory capital requirements. Furthermore, throughout bad times, the banking system may also experience further unexpected loan losses emanating from lower RRs. This is a major issue for reporting entitles, as in line with IFRS 9, they are required to provide forward-looking judgments.

Therefore, if RRs may be accurately forecasted during different cycles, capital buffers would be an effective tool to address the countercylicality nature of the economy, managing the complex environment of absorbing any unexpected credit losses. This view ensures the banking sector applies appropriate prudential practices, including maintaining capital requirements and controls for banks to operate within a bad cycle, but flexible enough to adjust accordingly during better periods. This has important implications for the pro-cyclicality effects of credit risk models, particularly the larger banks using an AIRB approach.

The remainder of our paper proceeds as follows. In Section \ref{Section 2}, we review the literature determining RRs and the proposed econometric specifications for the model calibrated with and without the latent credit cycle variable. Section 3 contains a description of the Moody's dataset used for the empirical analysis. The Bayesian inferential framework and corresponding Markov Chain Monte Carlo (MCMC) simulation algorithms are then detailed in Section \ref{Section 4}. This is followed by the results of our analysis and evaluation of our proposed models and their predictive capability in Section \ref{Section 5}. Section \ref{Section 6} concludes with a discussion and suggested directions for future research.

\section{Literature review} \label{Section 2}
Financial market participants, including bank regulators, are increasingly concerned with the management of risky assets, particularly bank loans. It is crucial to consider risk factors and market conditions at the time of placing an investment, but how these factors vary over time is also becoming increasingly viewed as critically important for making lending decisions. It is of particular importance to be mindful of the PD and subsequent loan recovery prospects during different or extreme conditions, as this will be critical to achieving expected return to finance providers, such as banks. The consequences of not considering this state- dependent risk can be severe for a bank, and may reach far beyond manageable levels, as was demonstrated in the financial crisis of 2008. This profound failure in prudential regulation, and corporate governance, attributable to poor operational risk management practices (\cite{financial2011financial}), underscores the importance of understanding financial risk, and in particular, credit risk, when pricing corporate loan contracts.

When a corporate borrower defaults, the lender endeavours to recover the outstanding debt using available collateral and liquidity mechanisms. Default alone, while not ideal for the borrower in terms of their ability to establish or maintain a strong credit rating, does not imply that the outstanding balance of the loan cannot ultimately be fully recovered during the post default period. In the vast majority of cases, lenders recover their full entitlement. However, there are many occasions when none or only a part of the outstanding indebted balance is recovered. Not surprisingly, the projected RR in the event of default is one of the key inputs when determining the price of any credit-related financial contract, including the value of the fundamental investment itself. Loss Given Default (LGD) is defined as one minus the RR, where the RR represents the proportion of the borrowed funds recovered (referred to as the exposure at default, or EAD) after the borrower goes into default. Hence, it is important for lenders to understand the factors that affect the actual RR, so appropriate decisions, including loan terms, can be made. Further motivation is provided by regulators, who require financial institutions to show evidence of prudential planning and modelling, and to ensure regulated capital requirements, as specified by the Basel III or governmental financial regulators, are maintained.

There are three main variables that determine the credit risk of a credit-based financial contract: the PD, the RR in the event of default and exposure at default (EAD), which is the total value to which the lending institution is exposed. \cite{altman2004default} points out that while significant attention has been paid to PD, RR and its apparent inverse relation with PD has attracted less attention. Notably, the RR is often treated as a constant variable, independent of PD. Existing studies have documented some empirical irregularities in the observed RR distribution. \cite{schuermann2004we} finds that the concept of average recovery, a quantity often reported by rating agencies, is potentially very misleading, as the recovery distribution is restricted to exist over the unit interval. This restriction implies that the lender cannot lose or recover more than the outstanding amount at the time of default. Notably, the observed RR distribution is typically U-shaped, with the largest relative frequency occurring near or at unity, a non-negligible mass around zero, and a spread of RRs observed across the interval itself. A flexible nonlinear model is more appropriate to reflect the relationship of this RR distribution with a number of loan or firm characteristics. 

Additionally, a range of econometric methods have been previously used to study RRs, including ordinary least squares (OLS), beta regression and a quasi-maximum likelihood estimation (QMLE) and a fractional regression methodology, as explored by \cite{gupton2002losscalctm}, \cite{acharya2007does} and \cite{khieu2012determinants}, respectively. Such analyses provide further insight into the possible determinants of a loan’s expected RR. Nevertheless, most approaches have some shortcomings. Notably, standard OLS ignores the unique distributional aspects of the observed RRs, despite the fact that the resulting RR values predicted from the model need not be bounded between zero and one. It also assumes constant marginal effects for each of the explanatory variables, a feature that is also unlikely given the constrained RR distribution. Furthermore, while the beta distributions underpinning a beta regression framework covers some variation of distributional densities over the unit interval, they cannot simultaneously accommodate a relative frequency mass in the middle of the unit interval with the relative large frequency masses observed around zero and one (\cite{de2004measuring}). Finally, while the model underlying the QMLE-based approach accommodates the constraints of the observed RR values, it does so at the expense of a coherent distribution model, fitting as it does a model for (binary) Bernoulli observations when in fact the values of RR may also lie inside the zero to one range, with clustering at 0 or 1.

The unsatisfactory features of the existing parametric approaches in the RR context have led to the development of more flexible models. Nonparametric methods have been shown to sometimes outperform their parametric counterparts in terms of accommodating non-linear relations between observed RRs and certain conditioning variables (\cite{qi2011comparison}). However, \cite{bastos2010forecasting} and \cite{qi2011comparison} find such flexible predictive models are more likely to over-fit the data and do not tend to work well in predicting future defaulted loan recoveries. Similarly, regression trees can become overly large and appear to produce results sensitive to assumed distribution and the dataset used. For example, \cite{bastos2010forecasting} and \cite{qi2011comparison} report very distinctive trees based on different datasets.

Modelling the determinants of RRs has shown them to be a function of individual loan characteristics, firm characteristics or fundamentals, industry variables, recovery process variables and macroeconomic factors. However, \cite{altman2005link} also demonstrates a negative association between an aggregate measure of the underlying default rate over a given period and the average RR, suggesting that changes in the underlying credit environment can also impact RRs. \cite{hu2002dependence} observe a similar negative relationship from data covering the period 1971-2000. In response, \cite{bruche2010recovery} define a two-state latent credit cycle variable and suggests that a 99\% credit VaR (the Value at Risk for a portfolio consisting of bank loans and corporate bonds) is underestimated by more than 1.5\% of the total outstanding amount if the credit cycle is omitted. The focus of this current research is to enhance the understanding of RR for corporate loans, addressing the limitations of the abovementioned literature. This is achieved by investigating RR in relation to appropriate firm and loan variables, in a time variant framework that allows for different economic circumstances, including major shocks such as the financial crisis of 2008. We use a less restrictive Bayesian, non-linear approach, accounting for time variation in the relationship between RR and PD.

\section{Data from Moody's Ultimate Recovery Database (1987--2015)}
We investigate RRs from Moody's Ultimate Recovery Database over the period 1987 through to 2015. A suitable RR variable is selected, with associate loan recovery processes and borrower characteristics extracted from the same dataset. In addition, we obtain macroeconomic and industry variables, such as a credit spread variable as a proxy for the PD, to be used as the RR determinants. The definitions, data sources and features of these determinants are provided in Section \ref{description}, following information about the RR variable. 

\subsection{Data description}\label{description}
The recovery data and other information about the defaulting firms and instruments are extracted from Moody's Ultimate Recovery Database, resulting in a set of 1,611 defaulted bank loans of US firms originated by an array of syndicated lending institutions over the period 1987 through 2015. Consistent with other empirical studies within the recovery literature (\cite{khieu2012determinants} and \cite{altman2014ultimate}), we use Moody's Discounted Recovery Rate variable as the relevant empirical RR measure.

A complete set of loan recovery determinants associated with each of the observed RRs, as used by \cite{khieu2012determinants}, is also employed. Broadly speaking, the available determinants address: loan characteristics, recovery process, borrower characteristics, as well as macroeconomic, industry and PD. The observed determinants relating to borrower characteristics are obtained through manual matching of these firms with Standard \& Poor’s Compustat firms based on both CUSIP numbers and also firm names. Definitions for each of these determinants are provided in Table \ref{tab:rrdets}. Typically, the defaulted loans have debt values, at the time of default, of greater than \$50 million.  This information, provided in Section \ref{determaints}, follows discussion of the key features of the RR data.

\subsection{Recovery rates and their determinants}\label{determaints}
The RR variable we use is defined as the nominal settlement recovery amount discounted back from each settlement instrument’s trading date to the last date cash was paid on the individual defaulted instrument, using the instrument’s own effective interest rate. This key recovery measure takes account of the time value of money for the effective settlement period. The sample here, covering 1987 through 2015, is split between term loans (48\%) and revolvers (52\%), i.e., loans that can be repaid and re-drawn any number of times within a term. About 7.6\% of the recoveries have some type of reorganization plan that shareholders have approved prior to, or at the time of, the bankruptcy filing.

\begin{figure*}[h!]
	\centering
	\includegraphics[scale=0.5]{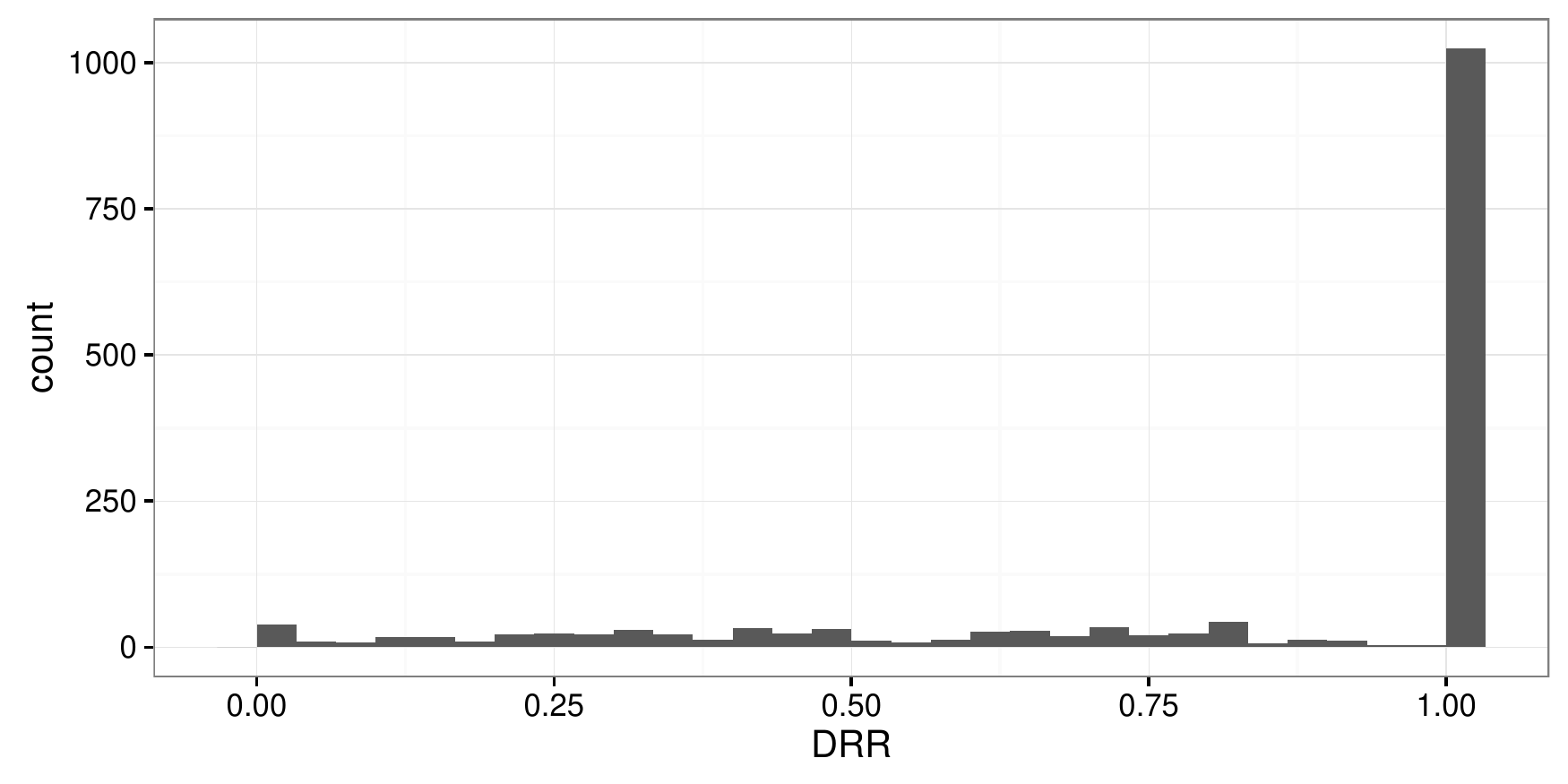}\includegraphics[scale=0.5]{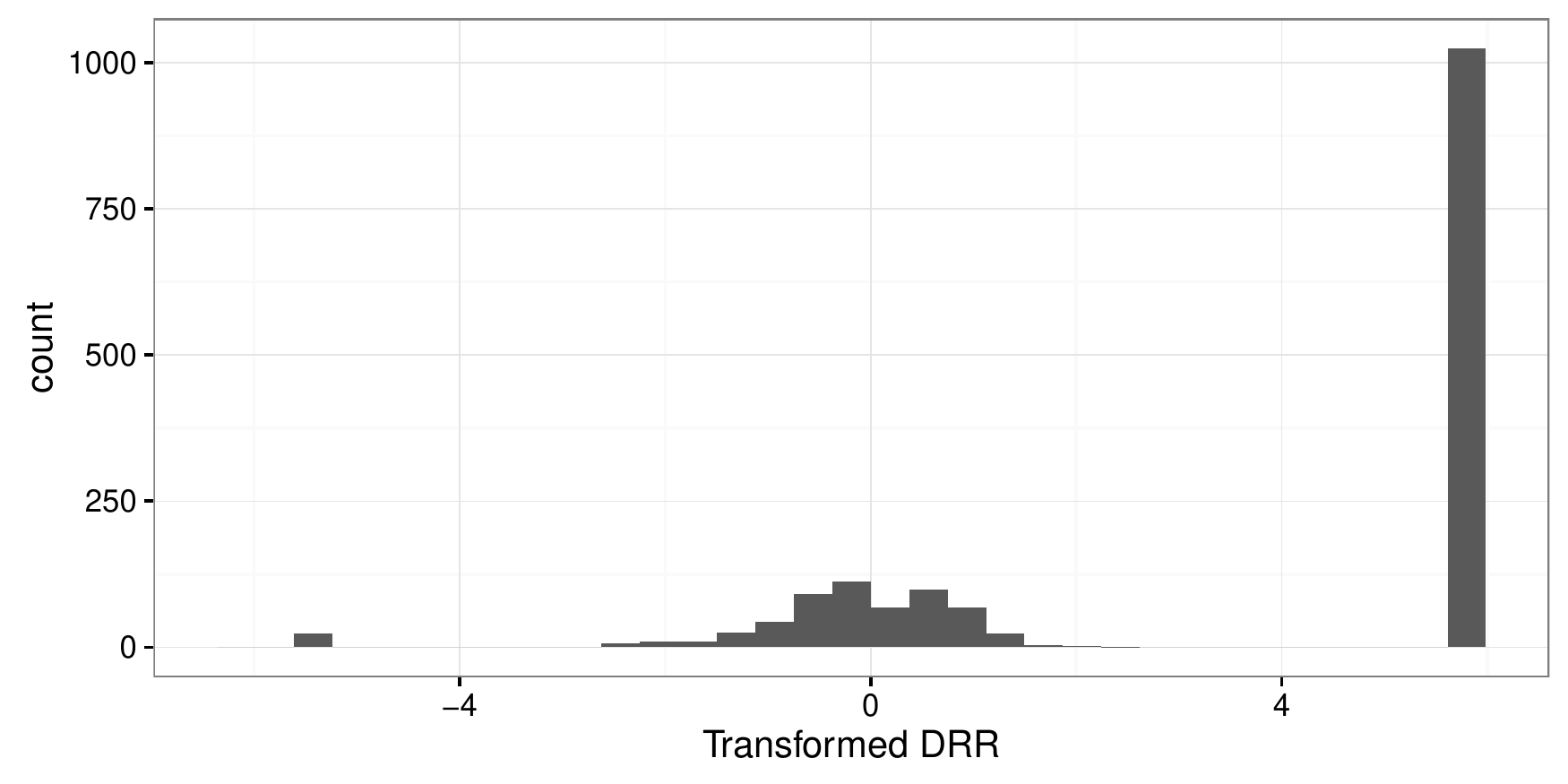}
	\caption{\small Histograms of the discounted recovery rates (RR) (left panel) and of the transformed discounted RRs ($\mathbf{y}$) (right panel), 1987-2015.}
	\label{fig:histogram}
\end{figure*}

Frequency plots (histograms) of both the raw RR and its transformed values (corresponding to (\ref{eq_yi})) for the sample period are shown in Figure \ref{fig:histogram}. Note the data concentration on the extreme right boundary of both graphs. While extreme modes associated with RR values at zero and unity are present, the mode at zero is almost negligible compared to that associated with full recovery. This feature is in line with the tendency towards left skewness typically associated with RRs for bank loans \cite{altman2014ultimate}. Given that the observed distribution of the RR is neither symmetric nor unimodal, the use of the average or median recovery as a single summary measure for the entire distribution is potentially very misleading. In particular, for this sample the average RR is 80.8\%, while the median value is 100\%, indicating that the distribution is indeed skewed to the left. It can be seen that both boundaries of the distribution have attracted concentrations, corresponding to the extremes of no recovery on the left and full recovery on the right, although the right boundary behavior dominates.

\begin{figure*}[h!]
	\centering
	\includegraphics[scale=0.5]{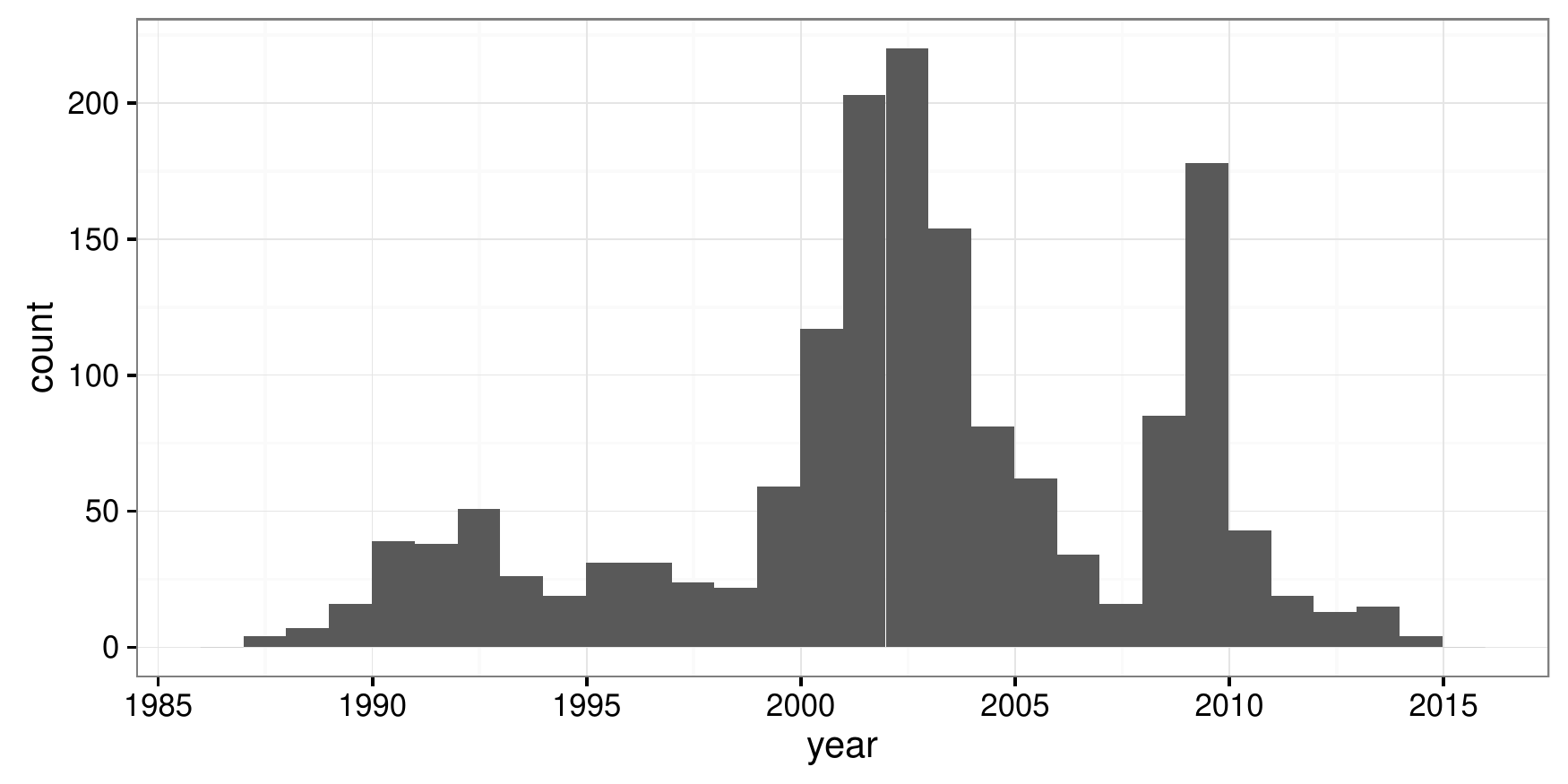}\includegraphics[scale=0.5]{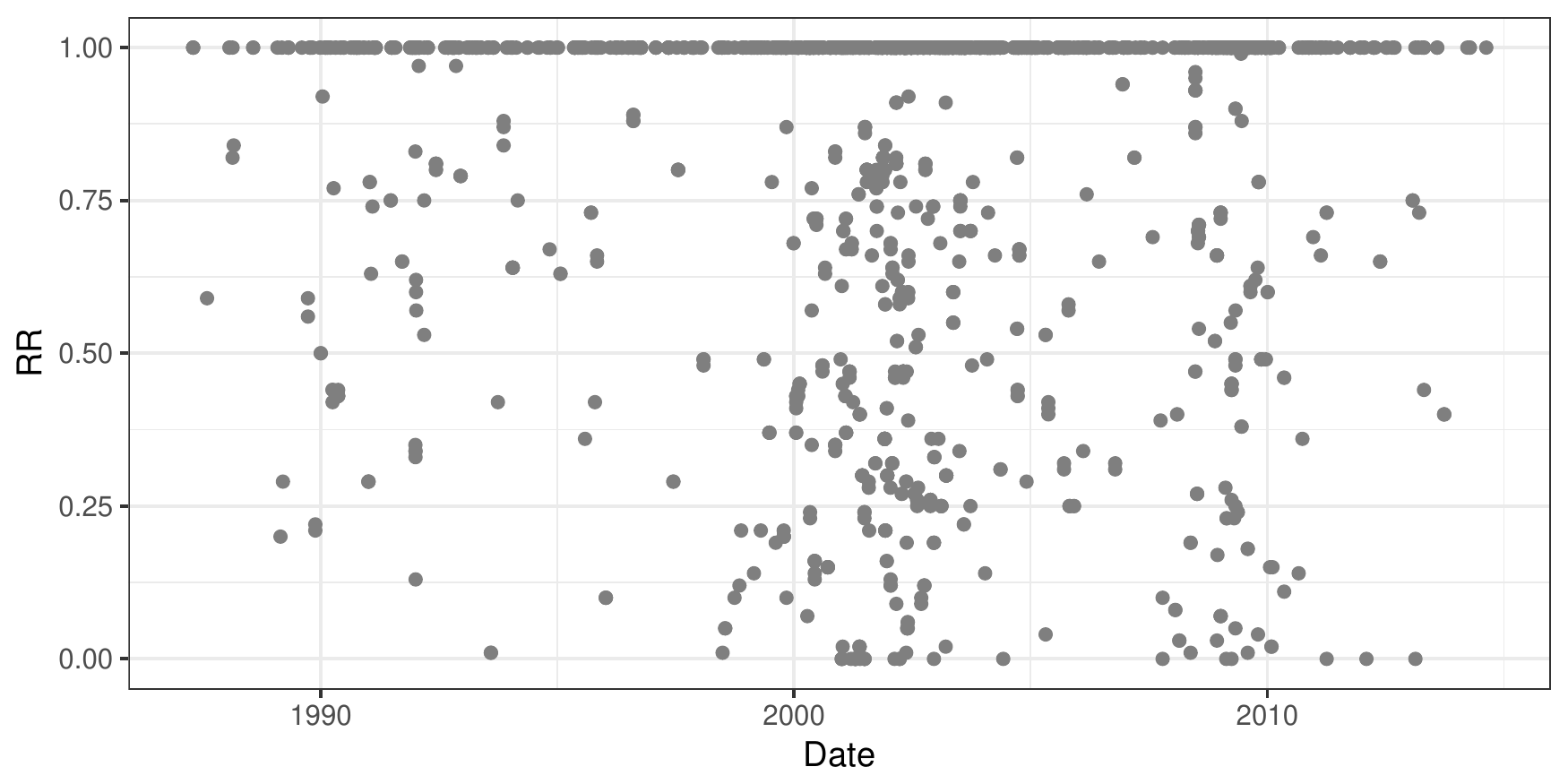}
	\caption{\small Number of defaults (left panel) and the RR outcomes by year (right panel), 1987-2015.}		
	\label{fig:databytime}
\end{figure*}

\begin{table*}[h]
\begin{center}
	\fontsize{7}{9}\selectfont
	\begin{tabular}{l*{8}{l}r}
	Variable & $\bm{n}$ & \textbf{Mean} & \textbf{Median} & \textbf{1st Qu} & \textbf{3rd Qu} & \textbf{Min}  & \textbf{Max} & \textbf{Binary} \\
	\hline
	RR & 1611 & 0.81 & 1.00 & 0.66 & 1.00 & 0.00 & 1.00 & N &\\
	\textbf{\textit{Loan characteristics}}&&&&&&&&&\\
	(1) LOANSIZE(\$M)              & 1611 & 224.5   & 96.0   & 35.0   &  208.5    & 1.0  & 11150.0  &  N  \\
	(2) LOANTYPE                   & 1611 & 0.48    & 0      & 0      &  1        & 0    & 1        &  Y  \\
	(3) LOANTYPE $\times$ FIRMSIZE         & 1611 & 813.5 & 0.0 & 0.0 & 692.3 & 0.0 & 60631.9 &  N  \\
	(4) ALLASSETCOLL               & 1611 & 0.62    & 1      & 0      &  1        & 0    & 1        &  Y  \\
	(5) INVENTRECIVECOLL           & 1611 & 0.10    & 0      & 0      &  0        & 0    & 1        &  Y  \\
	(6) OTHERCOLL                  & 1611 & 0.17    & 0      & 0      &  0        & 0    & 1        &  Y  \\
	\hline
	\textbf{\textit{Recovery process characteristics}}&&&&&&&&&\\
	(7) PREPACK                  & 1611 & 0.08    & 0      & 0      & 0      & 0    & 1           &  Y  \\
	(8) RESTRUCTURE              & 1611 & 0.13    & 0      & 0      & 0      & 0    & 1           &  Y  \\
	(9) OTHERDEFAULT             & 1611 & 0.01    & 0      & 0      & 0      & 0    & 1           &  Y  \\
	(10) TIMETOEMERGE             & 1611 & 13.49   & 9.67   & 2.59   & 18.42  & 0    & 156.33      &  N  \\
	(11) TIMETOEMERGE$^2$             & 1611 & 427.17  &  93.51 & 6.682   & 339.11  & 0.00    & 24439.07   &  N  \\
	(12) PREPACK $\times$ TIMETOEMERGE    & 1611 & 0.21  & 0.00 & 0.00 & 0.00  & 0.00    & 11.87  &  N  \\
	\hline
	\textbf{\textit{Borrower characteristics}}&&&&&&&&&\\
	(13) FIRMSIZE            & 1611 & 1654.8 & 665.5    & 227.4  & 1365.7 & 0      & 60631.92 &  N  \\
	(14) FIRMPPE             & 1611 & 0.54   & 0.44     & 0.13   & 0.82   & 0      & 9.73     &  N  \\
	(15) FIRMCF              & 1611 & 0.16   & 0.09     & 0.05   & 0.14   & 0      & 23.48    &  N  \\
	(16) FIRMLEV             & 1611 & 1.08   & 0.94     & 0.77   & 1.26   & 0      & 4.9      &  N  \\
	(17) EVERDEFAULTED       & 1611 & 0.15   & 0        & 0      & 0      & 0      & 1        &  Y  \\
	\hline
	\textbf{\textit{Macroeconomic and industry conditions}}&&&&&&&&&\\
	(18) GDP            & 1611 & 2.61   & 2.80     & 0.98   & 4.09   & 0.06   & 4.79     &  N  \\
	(19) INDDISTRESS    & 1611 & 0.18   & 0        & 0      & 0      & 0      & 1        &  Y  \\
	\hline
	\textbf{\textit{Probability of default}}&&&&&&&&&\\
	(20) AIS            & 1611 & 0.04   & 0.03     & 0.02   & 0.04   & 0      & 0.3      &  N  \\
	\hline
	\end{tabular}
		\captionof{table}{\small Descriptive statistics of the discounted RR and of the determinants of bank loan recoveries, by determinant category. By column corresponding to the variable indicated in the far left-hand column, the following statistics are reported: sample size ($n$), sample mean (Mean), sample median (Median), 25\% quantile (1st Qu), 75\% quantile (3rd Qu), sample minimum (Min), sample maximum (Max) and whether variable is binary [Y] or not [N] (Binary).} 
	\label{tab:descriptivestats}
\end{center} 
\end{table*}

Details of each of the RR determinants considered are described in Table \ref{tab:rrdets}. Additional descriptive statistics for the determinants are provided in Table \ref{tab:descriptivestats}, along with those for the observed RR data. Overall, the summary statistics of the RR determinants for the sample period here are similar to those found in \cite{emery2007moody} and \cite{khieu2012determinants}, suggesting that in aggregate the data here are, by and large, in line with those of earlier studies.  Furthermore, as per \cite{khieu2012determinants}, the firm characteristics in our baseline analyses are measured one year before default. In terms of loan characteristics, however, our dataset reports larger average loan sizes (\$224m compared to \$142m) than \cite{khieu2012determinants}, with similar increases in the average term loan size, in the average revolver value and in the values of loans secured by all assets. With respect to recovery process characteristics, the average length of time a sample firm stays in default is 13 months with a maximum of 13 years.  Most of the firms in the sample defaulted in a non-prepackaged bankruptcy, whereas 10\% went through prepackaged bankruptcy and 13\% had private workouts. Similar to \cite{mcconnell1996prepacks} and \cite{khieu2012determinants} the mean RR for loans with a reorganization plan lie between those for loans resolved in the traditional bankruptcy and those for loans going through the other forms of default resolution. With respect to borrower characteristics, the mean and median cash flows, relative to total assets for the settlement sample firm, are 16\% and 9\%, respectively.

	\section{A hierarchical econometric model for bank loan recovery rates} \label{Section 4}
	
	The starting point for our investigation is to determine the role of a complete set of recovery determinants, as explored by \citet{khieu2012determinants}. However in this research we implement a more enhanced and flexible modelling framework, and also use an extended sample that includes the GFC to ensure the capture of different economic conditions. The proposed methodology addresses two key challenges previously identified in the RR modelling literature, namely that the observed RR distribution has a distinctive (non-Gaussian) shape, and that when plotted over time the RRs appear to exhibit varying behavior - possibly owing to differences in the underlying PD. The first issue is addressed through the use of a finite Gaussian mixture model, first implemented on a combined loan and bond dataset by \citet{altman2014ultimate}. This approach enables the RR determinants to be stochastically connected to the observed RRs through a latent predictive regression structure. In addition, to capture cyclical aspects such as the impact of the GFC, we augment the model structure with a Markov switching mechanism within the predictive regression model. In this framework, the regression coefficients depend on the state of the credit cycle, where the state corresponds to either a `credit upturn' or a `credit downturn' - i.e. a `good' state or a `bad' one. The coefficient estimates we obtain for each credit state provide insight into the procyclical effects of RR determinants. To combine the Gaussian mixture components, the latent predictive regression and Markov switching mechanism, a hierarchical model is developed and estimated using a fully Bayesian inferential approach. The Bayesian approach enables a flexible hierarchical structure, which is estimated jointly and has the benefit of the consideration of each model component individually (i.e., marginally) while accounting for the uncertainty present in the remaining components. 
	
	Before detailing the form of the hierarchical model, also known as a `state space' model, and describing the associated Bayesian inferential framework, following \citet{altman2014ultimate} we transform the observed RRs from the unit interval to the real line via the inverse of the cumulative distribution function (cdf) associated with the standard normal distribution, denoted by $\Phi^{-1}(\cdot)$. Specifically, if $RR_{i}$ denotes the observed (appropriately discounted) RR value associated with defaulted loan $i$, we obtain the transformed RR value, denoted by $y_i$ and given by
	\begin{align}
	y_i = \Phi^{-1}(RR_i^\ast) \label{eq_yi}
	\end{align}
	where 
	\begin{align*}
	RR_i^\ast = \left\{ 
	\begin{array}{cl}
	\epsilon & \mbox{if } RR_i = 0\\
	RR_i & \mbox{if } 0 < RR_i < 1\\
	1-\epsilon & \mbox{if } RR_i = 1,
	\end{array} 
	\right.
	\end{align*}
	for $i=1,2,\ldots, n$. As is typical, before transformation is undertaken, the values of $RR_i$ at zero are replaced with a small positive value, $\epsilon$, and values at unity are replace with $1-\epsilon$, so that the $y_{i}$ values are all finite.\footnote{We use $\epsilon = 1 \times 10^{-8}$.} It is the distribution of these $y_{i}$ values that we model. Note that positive $y_i$ values result whenever the original $RR_{i}>0.5$. We now turn to the hierarchical model specification and the Bayesian inferential framework used to estimate it. Section \ref{Section 4.1} first details the Gaussian mixture model where membership to each mixture component is predicted by a latent regression on RR determinants. The regression coefficients here are shown in their static form, without the Markov switching component, which is described later in Section \ref{Section 4.2}. Section \ref{Section 4.3} then summarizes a computational strategy suitable for Bayesian inference to be conducted for the full dynamic model. Details regarding the algorithms required and implementation of the computational strategy are given in Section B of the Appendix.
	
	\subsection{A mixture model from recovery rate determinants}\label{Section 4.1}
    Having transformed each original RR observation, $y_i$ is then treated as arising from one of $J$ distinct Gaussian distributions, with the $j^{th}$ distribution having mean and variance denoted by $\mu_j$ and $\sigma^2_j$, with $\mu_1 < \mu_2 < \cdots <\mu_J$. From the investor's perspective, recovery outcomes from a mixture component having a larger mean will be preferred, e.g. the $J^{th}$ mixture component is preferred over the $(J-1)^{st}$, etcetera, with the first component being least desired, and therefore the ordering is imposed to retain the ability to interpret each of the categories.
		
   Next, the connection between the mixture components and the RR determinants occurs through a latent ordered probit regression framework (\cite{albert1993bayesian}), which permits a range of explanatory variables,  including loan, borrower, recovery process and macroeconomic or industry conditions, to characterize the probability of $y_{i}$ being in component $j$ of the Gaussian mixture. In particular, the determinants associated with loan $i$, denoted by $x_{1,i},x_{2,i},\ldots,x_{K,i}$, are related to a latent variable $z_i$ through the regression equation
	\begin{align}
	z_i = \beta_0+\beta_1x_{1,i}+\cdots+\beta_Kx_{K,i}+\varepsilon_i, \label{Eq_CF2}
	\end{align}
	with $\varepsilon_i \sim N(0,1)$. The latent (unobserved) $z_i$ is referred to as a \emph{predictive score} for defaulted loan $i$, while the vector $\boldsymbol{\beta}=(\beta_0,\beta_1,\ldots,\beta_K)'$ contains the regression coefficients that describe the marginal impact of each of the determinants on this predictive score. 
	The predictive score for loan $i$ in (\ref{Eq_CF2}) relates to each of the $J$ Gaussian mixture components via a set of so-called `cut-points', $\mathbf{c}=(c_0,\cdots,c_J)$, with $c_0= -\infty, c_1=0, c_J=+\infty,$ so that when in fact loan $i$ belongs to group $j$, the $j^{th}$ mixture probability may be calculated as $\Pr(c_{j-1} < z_i \leq c_{j}).$ Although the values of $c_0, c_1$ and $c_J$ are fixed for identification purposes (see again \citet{albert1993bayesian}) the locations of the remaining cut-points (here $c_2$ and $c_3$) are treated as unknowns to be estimated. The values of $\mu_j$ and $\sigma^2_j$ are also estimated, essentially being determined by those $y_i$ that are predicted by the regression to fall in category $j$. 	
	
	Up to this point, the approach used here largely follows that of \citet{altman2014ultimate}, apart from our use of a wider set of determinants as discussed in Section \ref{section_intro}. However, as described the next section, we introduce an additional Markov switching component to the framework, so that the impact of the RR determinants is able to vary with the credit environment, whether good or bad, at the time of default. We also introduce a prior for $\boldsymbol{\beta}$, associated with the LASSO and discussed in Section \ref{sec:PriorLasso}.
	
	\subsection{The credit cycle}\label{Section 4.2}
	To incorporate the notion of an underlying dynamic credit cycle, a two-state Markov switching component is added to the mixture model with latent predictive regression framework outlined in Section \ref{Section 4.1}. This binary credit cycle state variable for time (year) $t$ is denoted by $S_t$, and takes on either the value of zero or one, depending on the underlying credit environment prevalent at the time of default.\footnote{In this study, $t=1$ corresponds to 1987, the first year of the available sample period.}  The credit cycle states are normalized so that $S_{t} = 0$ corresponds to a low recovery period (a downturn, or `bad' credit state), while $S_{t} = 1$ corresponds to a high recovery period (an upturn, or `good' credit state). Transition to each credit cycle state at time $t$ from a relevant state one period earlier, time $t-1$, is governed by the probabilities
	\begin{align}\label{MSprob}
	\begin{split}
	& \mbox{Prob}(S_t = 0|S_{t-1} = 0) = p \\
	& \mbox{Prob}(S_t = 1|S_{t-1} = 0) = 1 - p \\ 
	& \mbox{Prob}(S_t = 1|S_{t-1} = 1) = q \\ 
	& \mbox{Prob}(S_t = 0|S_{t-1} = 1) = 1 - q.
	\end{split}
	\end{align}
	
	According to the transition probabilities in (\ref{MSprob}), if the credit cycle at time $t-1$ is in a low recovery state (i.e $S_{t-1}=0$), then the chance of remaining in this `bad' state at time $t$ equal to $p$, with $0<p<1$, while the chance of moving to the `good' recovery state (i.e.with $S_t=1$) is equal to $1-p$. On the other hand, if $S_{t-1}=1$, then the chance of remaining in the `good' recovery state at time $t$ is equal to $q$, with $0<q<1$, and the chance of moving to the `bad' recovery state at time $t$ is given by $1-q$. 
	
	Using the Markov switching device, two sets of regression coefficients are obtained for the recovery determinants: $\boldsymbol{\beta_0} = (\beta_{0,0},\beta_{1,0},\ldots,\beta_{K,0})'$ relating to the predictive scores in credit cycle downturns, and $\boldsymbol{\beta_1}=(\beta_{0,1},\ldots, $ \\ $\beta_{K,1})'$ applicable during credit cycle upturns. 
	To link the latent credit cycle states to the available data, let $t_i$ denote the time associated with the default of loan $i$, so that $S_{t_i}$ indicates the relevant state of the credit cycle at the time of default of loan $i$. The predictive regression coefficient vector $\boldsymbol{\beta_0}$ will apply for predicting $z_i$ if $S_{t_i}=0$, whereas the vector $\boldsymbol{\beta_1}$ will apply for predicting $z_i$ if $S_{t_i}=1$. Hence, by adding the Markov switching component, the regression coefficients in (\ref*{Eq_CF2}) become state dependent, and the predictive regression for loan $i$ becomes
	\begin{align}
	z_{i} = \beta_{0,S_{t_i}}+\beta_{1,S_{t_i}}x_{1,i}+\cdots+\beta_{K,S_{t_i}}x_{K,i}+\varepsilon_i,
	\label{Eq_CF2switch}
	\end{align}
	where again $\varepsilon_i\overset{iid}{\sim}  \mathcal{N}(0,1)$, for $i=1,2,\ldots,n$. Now that the regression coefficients in the predictive regression are state dependent, the estimated values of the vectors $\boldsymbol{\beta_1}$ and $\boldsymbol{\beta_0}$ will provide insight into the differentiated impact of RR determinants in `good' times and `bad'. 

	\subsection{Bayesian inference}\label{Section 4.3}
    Like \cite{altman2014ultimate}, we take a Bayesian approach when estimating the proposed model, an approach that offers several advantages over the perhaps more familiar frequentist strategy. The outcome of any Bayesian inferential procedure is a full joint probability distribution for all unknowns, including both parameters and latent variables. This outcome distribution, referred to the joint posterior distribution, characterizes all that is known about the parameters, and the credit states, prediction scores and Gaussian mixture allocations for each loan. From this joint posterior, the corresponding marginal distribution for any individual parameter or state variable (or indeed any subset of these) will automatically and coherently account for uncertainty in the remaining unknown variables. This is of particular importance when working with a hierarchical model, such as the one we advocate here.
	
	An added advantage of using a hierarchical model within a Bayesian framework is that computation to produce the posterior can be undertaken efficiently using Markov chain Monte Carlo (MCMC) techniques. Details of this computation is provided in Appendix \ref{Appendix A}. As a further advantage, Bayesian inference yields a finite sample analysis, conditioning only on the available data, whereas a corresponding Frequentist inferential method would typically require assumptions about the behavior of estimators as the sample size increases without bound. This is important in empirical applications, such as the one undertaken here, where the number of RR observations are limited relative to the number of unknowns being estimated. 
	
	\subsubsection{Prior distribution incorporating the Bayesian LASSO } \label{sec:PriorLasso}
	We conservatively adopt a relatively non-informative prior distribution with \textit{a priori} independence assumed between $\boldsymbol{\mu}=(\mu_1, \mu_2, \ldots, \mu_J)'$, $\boldsymbol{\sigma^2}=(\sigma_1^2, \sigma_2^2, \ldots, \sigma_J^2)'$, $\mathbf{c},$  $ \boldsymbol{\beta_0}, \boldsymbol{\beta_1}, p$ and $q$. Apart from the prior specified for the state-dependent predictive regression coefficients, $\boldsymbol{\beta_0}$ and $\boldsymbol{\beta_1}$ discussed below, the prior components are chosen from the appropriate conditionally conjugate family, thereby enabling fast computation of the posterior distribution via MCMC. 
	
	For the predictive regression vectors, $\boldsymbol{\beta_0}$ and $\boldsymbol{\beta_1}$, we introduce the use of the Bayesian LASSO prior of \citet{park2008bayesian}. As is now widely recognized the LASSO encourages a sparse regression model by down-weighting certain covariates when a large number of regression terms are used (\citet{nazemi2017macroeconomic}), as is the case here. Effectively, the LASSO will reduce the size of the estimated regression coefficients to account for correlation \\(multi-collinearity) or other dependence between the available recovery determinants, favouring putting weight on regressors whose association with the response variable (here the latent predictive scores $\mathbf{z}=(z_1,z_2, \ldots, z_n)$) can be estimated with relative certainty.  In this way, predictive information shared by different determinants is not `double counted' when fitting the model. The Bayesian LASSO achieves this reduction, or \emph{shrinkage}, through the choice of the prior distribution for $\boldsymbol{\beta_0} \ \mbox{and} \ \boldsymbol{\beta_1}$. This prior distribution for each regression vector in the dynamic credit cycle context relies critically on certain additional so-called shrinkage parameters, denoted by $\lambda_0^2$ and $\lambda_1^2$, respectively, with a single shrinkage parameter, denoted by $\lambda^2$ used for the static latent regression model. These shrinkage parameters are included as unknowns, and are also estimated here within the Bayesian framework.
	
	We note that many existing studies have considered the predictive performance for RRs. For instance, \cite{altman2014ultimate} investigate the predictive performance of different models using a set of variables for debt seniority, collateralization and industry classification. In the Bayesian paradigm, \cite{barbieri2004optimal} point out that a model with highest posterior probability is not necessarily optimal for prediction, instead, optimal predictive models are `median probability models'.
	
	\section{Empirical results} \label{Section 5}
	The results reported in this section are based on two implementations of the model described in Section \ref{Section 4}, namely the \emph{static} version, corresponding to Section \ref{Section 4.2} where the predictive regression coefficient variables are assumed to be constant over the entire sample period, and the \emph{dynamic} version described in Section \ref{Section 4.3}, where the latent time-varying credit cycle is included. The LASSO priors are used in both cases. Note that the where the term estimate is used it will generally refer to the posterior mean of the posterior distribution for the relevant quantity. Uncertainty in such an estimate will be indicated by a 95\% so-called highest posterior density (HPD) interval taken as the shortest single interval associated with 95\% marginal posterior probability. These Bayesian point and interval estimates are used to summarize the marginal posterior distributions, and are obtained from the MCMC output based on 100,000 MCMC draws retained following a 5,000 burn-in period. 
	
	\subsection{Recovery Mixture components}	\label{5.1}
	
	As alluded to in Section \ref{description}, given RR observations are clustered at zero (zero recovery) and one (full recovery), following \cite{altman2014ultimate}, we apply a $J=4$ Gaussian mixture model to transformed RRs. Table \ref{tab:mixturecomps} provides details of the features of the estimated Gaussian mixture components that result from each of the two models fitted to the dataset considered. For each case, the estimated mean and standard deviation parameters for each Gaussian component are provided, along with its corresponding mixture weight and median RR. The estimated components labeled $1$ and $4$ effectively concentrate, with the same relative proportions for both the dynamic and static specifications, on point masses corresponding to RR values at zero and one, respectively. This fact confirms that the mixture specification accommodates the corresponding observed concentrations at the extremes found in the empirical RR distribution. However, the two interior mixture components (labeled $2$ and $3$) show differences across these attributes, notably in the third mixture component. As the two models correspond to different latent predictive regression structures - one static (i.e. without imposing the Markov switching credit states) and the other dynamic - two separate estimation results are shown. The mean parameter for the $j^{th}$ component, $\mu_j$, and the corresponding standard deviation, $\sigma_j^2$, is determined by observed RRs with predictive regressions correspond to outcomes that fall in mixture component $j$.
		
	\begin{table*}[h!]
		\begin{center}
			\fontsize{6}{6}\selectfont
			\begin{tabular}{|p{2.5cm}|c c c c | c c c c|}
\hline & \multicolumn{4}{c|}{\footnotesize\textbf{ }} & \multicolumn{4}{c|}{\footnotesize\textbf{ }} \\
				& \multicolumn{4}{c|}{\footnotesize\textbf{Static model}} & \multicolumn{4}{c|}{\footnotesize\textbf{Dynamic model}} \\
				\textbf{Component ($j$)} & \textit{1}& \textit{2} & \textit{3} & \textit{4}  & \textit{1}& \textit{2} & \textit{3} & \textit{4}\\
				\hline
				& &   &  &  & &  &   & \\
				a) Mean ($\mu_j$)   & -5.61 & -0.85   & 0.21   & 5.61  & -5.61 &  -1.31  & 0.09   & 5.61 \\
				& \scriptsize(-5.72,-5.50) & \scriptsize(-1.77,-0.32) & \scriptsize(0.03,0.46) & \scriptsize(5.61,5.61) & \scriptsize(-5.72,-5.50) & \scriptsize(-2.00,-0.52) & \scriptsize(-0.02,0.24) & \scriptsize(5.61,5.61) \\
				&&&&&&&&
				\\
				b) Std ($\sigma_j$)  & 0.08  & 0.54    & 0.41   & 0.00 &  0.08 &  0.44   & 0.48   & 0.00  \\
				& \scriptsize(0.05,0.13) & \scriptsize(0.21,0.83) & \scriptsize(0.29,0.53) & \scriptsize(0.00,0.00) & \scriptsize(0.05,0.13) & \scriptsize(0.14,0.90) & \scriptsize(0.38,0.58) & \scriptsize(0.00,0.00) \\
				&&&&&&&&
				\\
				c) Implied weight   & 0.01  & 0.10    & 0.25   & 0.64   &  0.01 &  0.04   & 0.31   & 0.64 \\
				& \scriptsize(0.01,0.01) & \scriptsize(0.02,0.20) & \scriptsize(0.15,0.33) & \scriptsize(0.64,0.64) & \scriptsize(0.01,0.01) & \scriptsize(0.01,0.11) & \scriptsize(0.24,0.34) & \scriptsize(0.64,0.64) \\
				&&&&&&&&
				\\
				d) Mean RR & 0.00  & 0.25    & 0.63   & 1.00 &  0.00 &  0.10   & 0.54   & 1.00\\
				& \scriptsize(0.00,0.00) & \scriptsize(0.04,0.37) & \scriptsize(0.51,0.68) & \scriptsize(1.00,1.00) & \scriptsize(0.00,0.00) & \scriptsize(0.02,0.30) & \scriptsize(0.49,0.59) & \scriptsize(1.00,1.00) \\
				&&&&&&&&
				\\
\hline
			\end{tabular}
		\end{center}
		\captionof{table}{\small Estimated Gaussian mixture components. Posterior mean and 95\% HPD interval (in parentheses) for each mixture component as indicated in the first row, and for a) the component mean parameter (Mean ($\mu_j$)) in row two; b) the component standard deviation parameter (Std ($\sigma_j$)); c) the Implied weight, as given by the proportion of observations allocated to the mixture component; and d) Mean RR, corresponding to the inversion of (\ref{eq_yi}); 
				for each of the four mixture components as labeled by $j=1, 2, 3,$ and $4$, for the static model (columns 2-5)  and the dynamic model (columns 6-9)}
		\label{tab:mixturecomps}
	\end{table*}

	\subsection{The latent credit cycle}\label{5.2}	
	
\begin{figure}[h!]
		\centering
		\includegraphics[scale=0.7]{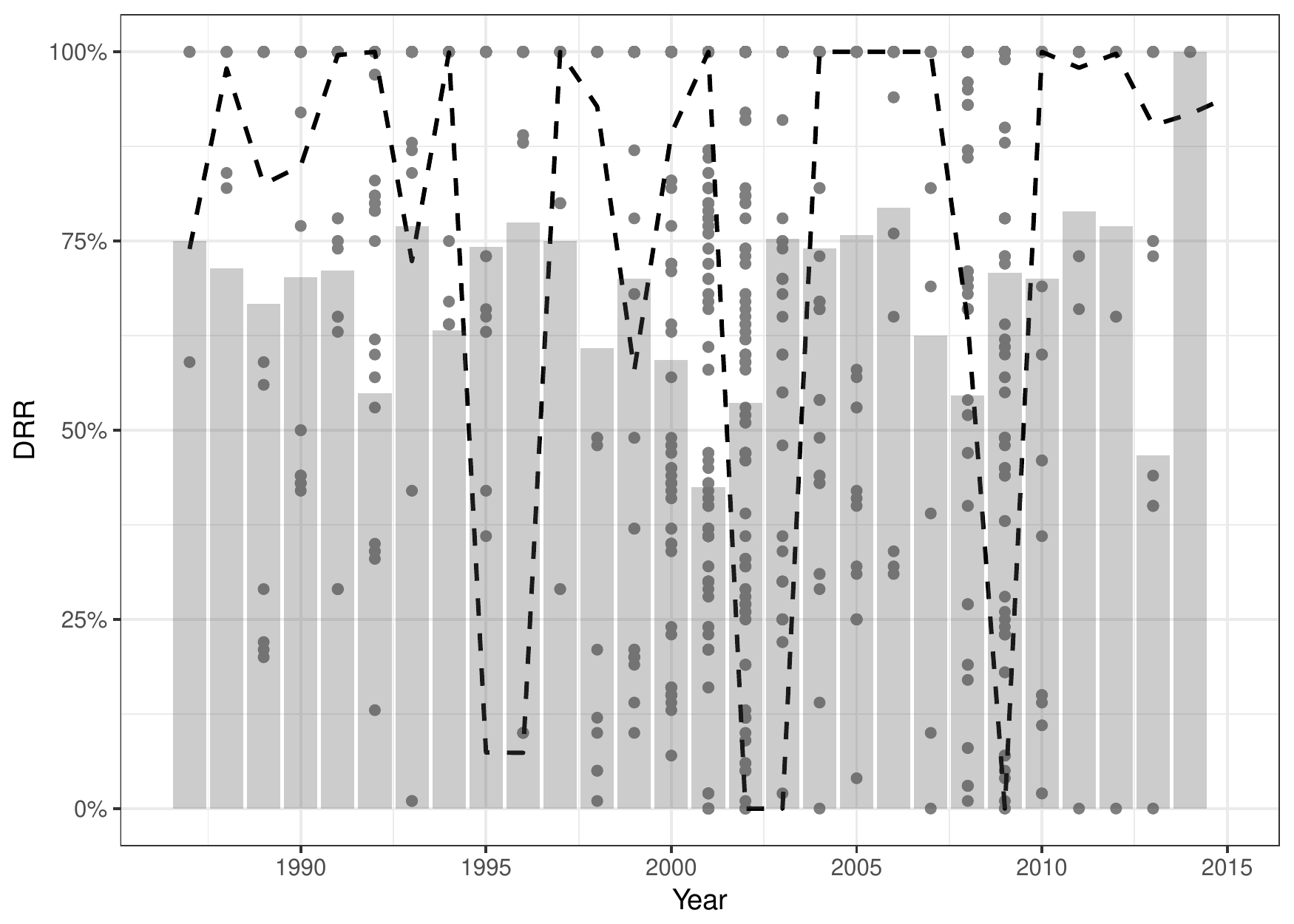}
		\caption{\small{RRs plotted over the 1987-2015 period and aligned by calender year, with the estimated probability of being in a `good' credit state, implied by the dynamic model, given by the superimposed line graph. Shading area represents the percentage of full recovery in each corresponding year.}} 
		\label{creditcycle}
	\end{figure}

	The latent Markov switching states are introduced into the dynamic model to characterize the time-series variation in the observed RRs. 
	Estimates of $q$, the probability of remaining in a good credit state, from one year to the next, and $p$, the probability of the economy remaining in a bad credit state, are given in Table \ref{tab:statesm4}, with the corresponding estimated steady state (or long-run) probabilities being 61\% and 39\% for the `good' and `bad' states, respectively, as indicated by the final row of Table \ref{tab:statesm4}. A line graph of the estimated  probabilities for being in the good credit cycle state during each specific year during the given sample period is overlaid on a plot of RR outcomes in Figure \ref{creditcycle}, with the shaded bars shown in the figure indicating the sample proportion of fully recovered RRs reported in each calendar year. Interestingly, the troughs that appear in the estimated credit cycle reflect, at least, two well known economic downturns, namely, the burst of US dot-com bubble in 2002 and global financial crisis (GFC) in 2008-2009. The indicated credit downturn corresponding to 1994-1995 may bear some connection to the Mexican peso crisis and its impact on the North American Free Trade Agreement.

	\begin{table}
		\begin{center}
			\fontsize{8}{7}\selectfont
			\begin{tabular}{|c|c c|}
				\hline
					\multirow{2}{*}{} & \multicolumn{2}{c|}{ } \\
				\multirow{2}{*}{} & \multicolumn{2}{c|}{$\Pr(S_t|S_{t-1})$} \\
				& $S_t = 0$ & $S_t = 1$ \\
				\hline
				\multirow{2}{*}{ }  & { } & { } \\
				\multirow{2}{*}{$S_{t-1} = 0$}  & 0.53 & 0.47 \\
				& \scriptsize(0.24,0.83) & \scriptsize(0.17,0.76) \\
					\multirow{2}{*}{ }  & { } & { } \\
				\multirow{2}{*}{$S_{t-1} = 1$}  & 0.67  & 0.33 \\
				& \scriptsize(0.40,0.88) & \scriptsize(0.12,0.60) \\
				\hline
					\multirow{2}{*}{ }  & { } & { } \\
				\multirow{2}{*}{Steady state: $\Pr(S_t=1)$} & \multicolumn{2}{c|}{0.61} \\
				&\multicolumn{2}{c|}{\scriptsize(0.38,0.83)}\\
				\hline
			\end{tabular}
			\captionof{table}{\small Estimated posterior mean and $95\%$ HDP (in parenthesis) for each possible transition probability associated with a one period transition from credit state $S_{t-1}$, shown by row in the first column, to a new credit state $S_t$, shown by corresponding entry in columns 2 and 3. The final row provides the estimated overall long-run probability of being in the `good' credit state $\Pr(S_t=1)$ resulting from the dynamic model.}
			\label{tab:statesm4}
		\end{center}
	\end{table}
	
	\subsection{The predictive regressions} \label{5.3}

		\begin{figure}[h]
			\centering
			\includegraphics[scale=0.65]{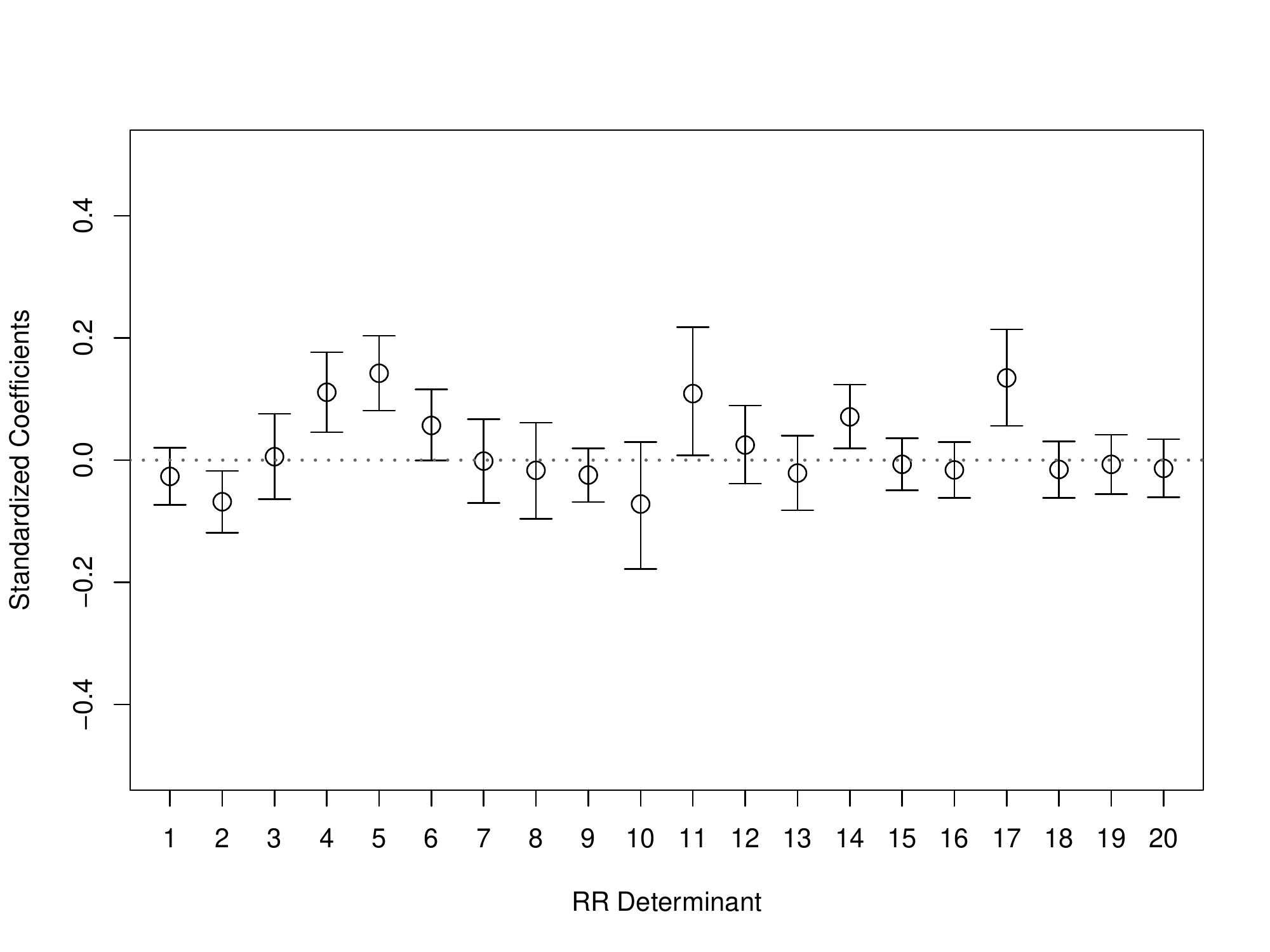}	
			\caption{Static model: Posterior mean estimates $(\circ)$ of individual $\beta_k$ coefficients, for variable $k=1,2,...,K=20$, with corresponding 95\% HPDs indicated by the vertical bars. The variables are: $(1)$ LOANSIZE, $(2)$ LOANTYPE, $(3)$ LOANTYPE $\times$ FIRMSIZE, $(4)$ ALLASSETCOLL,  $(5)$ INVENTRECIVECOLL, $(6)$ OTHERCOLL, $(7)$ PREPACK, $(8)$ RESTRUCTURE, $(9)$ OTHERDEFAULT, $(10)$ TIMETOEMERGE, $(11)$ TIMETOEMERGE$^2$, $(12)$ PREPACK $\times$ TIMETOEMERGE, $(13)$ FIRMSIZE, $(14)$ FIRMPPE, $(15)$ FIRMCF, $(16)$ FIRMLEV, $(17)$ EVERDEFAULTED, $(18)$ INDUSTRESS, $(19)$ GDP, $(20)$ AIS.}
			\label{fig:parametersm2}
		\end{figure}

		\begin{figure}[h!]
			\centering
			\includegraphics[scale=0.65]{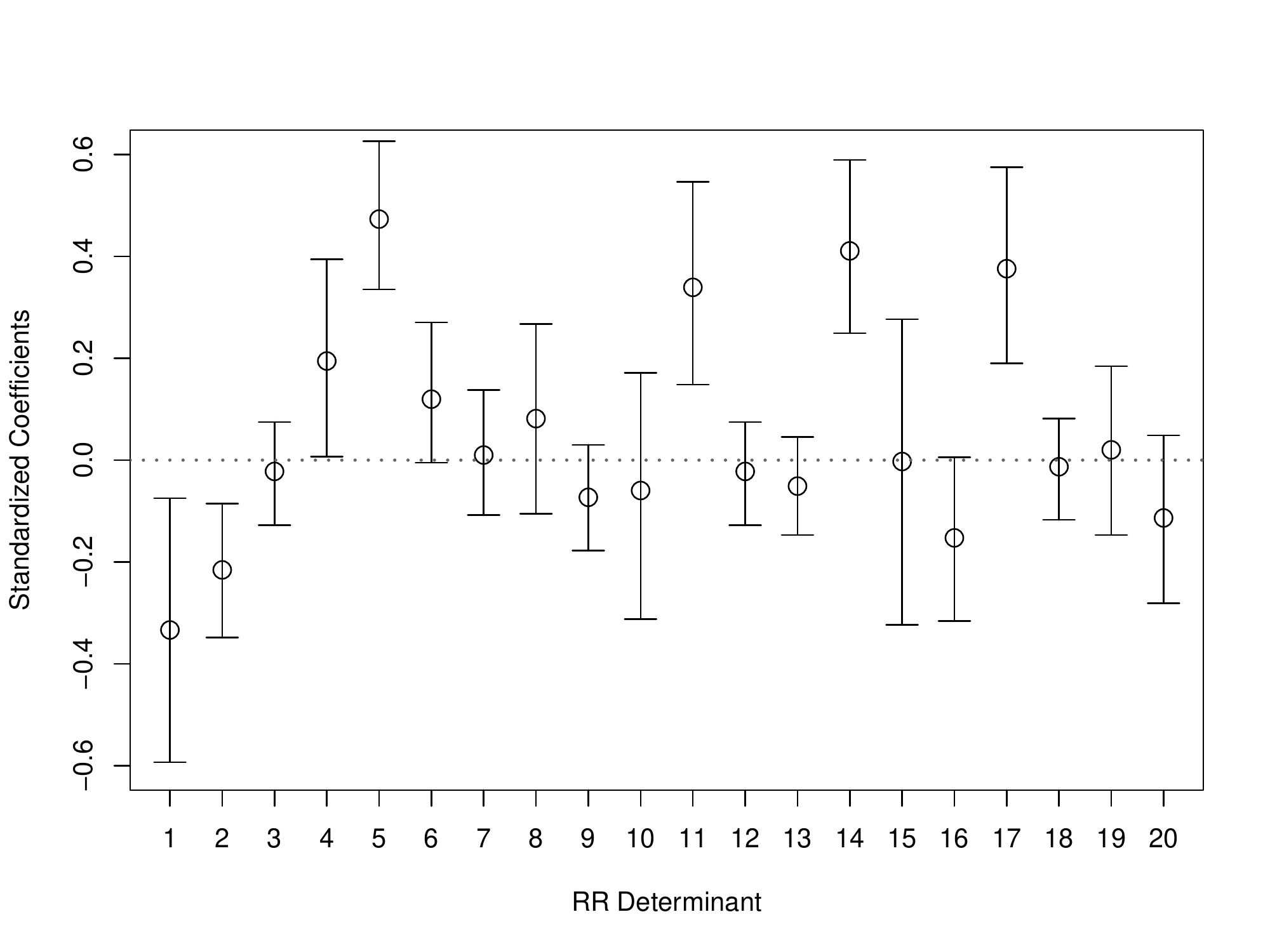}	
			\vspace*{\floatsep}
			\centering
			\includegraphics[scale=0.65]{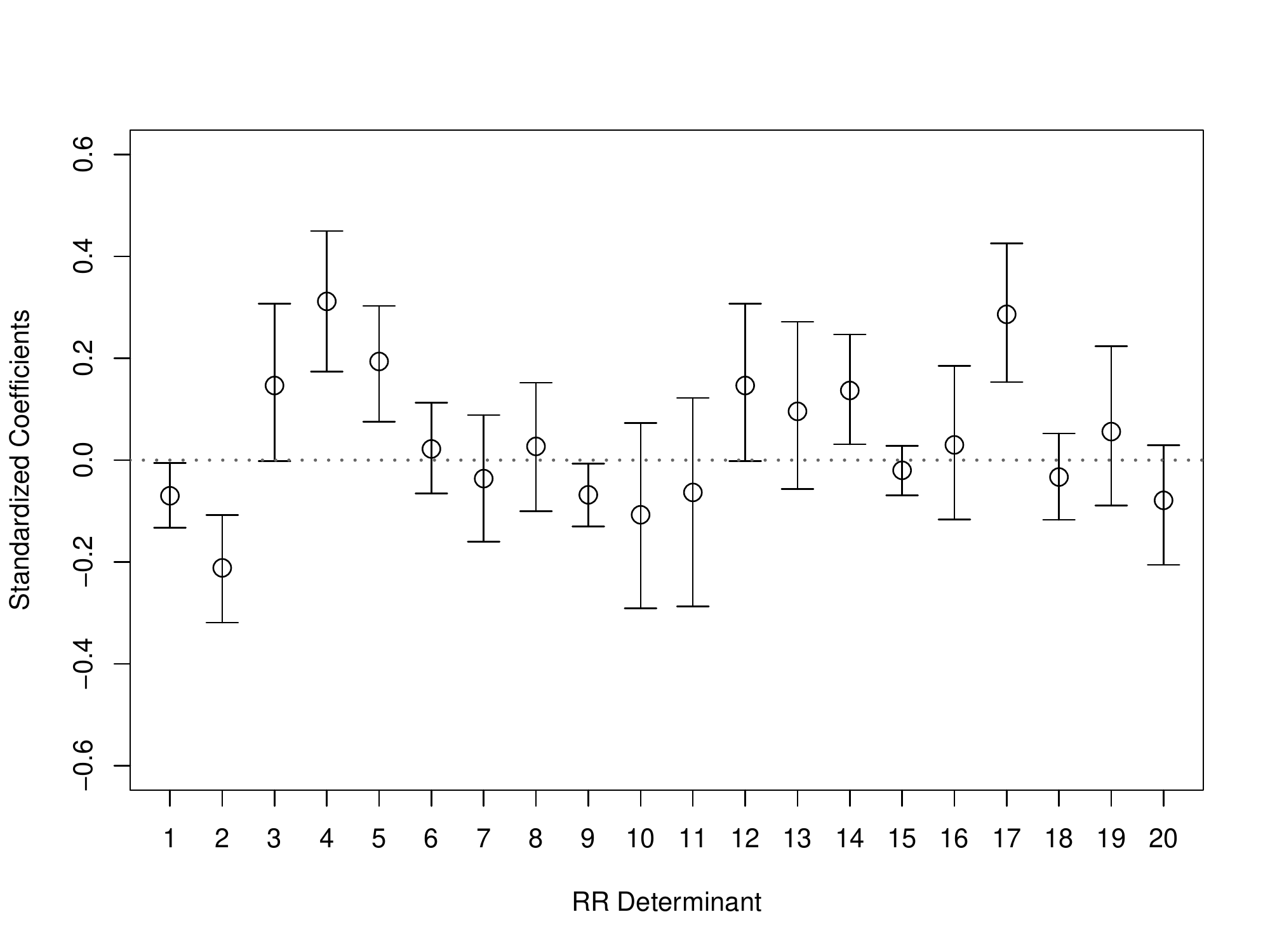}
			\caption{Dynamic model: Posterior mean estimates $(\circ)$ of $\beta_{0,k}$ (top panel) and $\beta_{1,k}$ (bottom panel), for variable $k=1,2,...,K=20$, along with corresponding 95\% credible intervals indicated by the vertical bars. The variables are: $(1)$ LOANSIZE, $(2)$ LOANTYPE, $(3)$ LOANTYPE $\times$ FIRMSIZE, $(4)$ ALLASSETCOLL,  $(5)$ INVENTRECIVECOLL, $(6)$ OTHERCOLL, $(7)$ PREPACK, $(8)$ RESTRUCTURE, $(9)$ OTHERDEFAULT, $(10)$ TIMETOEMERGE, $(11)$ TIMETOEMERGE$^2$, $(12)$ PREPACK $\times$ TIMETOEMERGE, $(13)$ FIRMSIZE, $(14)$ FIRMPPE, $(15)$ FIRMCF, $(16)$ FIRMLEV, $(17)$ EVERDEFAULTED, $(18)$ INDUSTRESS, $(19)$ GDP, $(20)$ AIS.}
			\label{fig:parametersm4}
		\end{figure}
		
	To demonstrate the importance of allowing for temporal variation in economic conditions, we contrast the inferential results from the dynamic and static Bayesian models. As discussed in Section 4, the models are developed using a Bayesian LASSO to control for multi-collinearity arising from competing and highly correlated RR determinants. Figures \ref{fig:parametersm2} and \ref{fig:parametersm4} illustrate the significance of each of the variables after applying the Bayesian LASSO, with Figure \ref{fig:parametersm2} showing the signficance of parameters in the static model, and Figure \ref{fig:parametersm4} showing those for the dynamic case, with the top panel of the latter figure corresponding to significance for the bad' credit state and the lower panel corresponding to the significance of determinants under the `good' credit state. In all cases, interval estimates for variables that cross the vertical axis at zero indicate a lack of (marginal) significance for that variable in the relevant model.

	In Table \ref{compareRRdets} we report an alternative summary of these predictive regression estimation results, again for each of the static and dynamic models, in this case showing the sign only of the significant coefficients along with those obtained previously in \cite{khieu2012determinants}. In column two, we report the signs of the significant RR determinants identified in column one under our static model resulting from the Bayesian approach and corresponding to data from 1987-2015. Columns three and four of Table \ref{compareRRdets} report the sign of significant RR determinants under the Bayesian dynamic model, with $\beta_0$ corresponding to the bad credit state (i.e. when $S_t = 0$) and with $\beta_1$ corresponding to the good credit states (i.e. when $S_t = 1$). For comparative purposes, the sign of the significant coefficients of these determinants corresponding to Frequentist inference using OLS and QMLE methodologies, and relating to data from 1997-2007 (as reported in \cite{khieu2012determinants}), are provided in columns five and six. This is done to illustrate the contribution made by static vs. dynamic versions, and the need to allow for variation in the impact of RR determinants under different credit conditions. The reported models are more parsimonious relative to the existing literature due to our use of a LASSO, though both are consistent overall regarding the relevance of RR determinants. While the numerical values of the estimated Frequentist and Bayesian coefficients themselves are not directly comparable, we can compare their statistical significance and their sign.
	
	\begin{table*}[h]
		\fontsize{7}{6}\selectfont
		\begin{tabularx}{\textwidth}{|m{4.8cm}| >{\centering\arraybackslash}m{2cm}| >{\centering\arraybackslash}m{2cm}| >{\centering\arraybackslash}m{2cm}| >{\centering\arraybackslash}m{2cm}| >{\centering\arraybackslash}m{2.05cm}|}
\hline
	& { } &\multicolumn{2}{c|}{ }& { } & { }\\ 
		& \textbf{Static model} &\multicolumn{2}{c|}{\textbf{Dynamic model}}& \textbf{\citet{khieu2012determinants}} & \textbf{\citet{khieu2012determinants}}\\ 
			&  $\beta$   & $\beta_0$  &$\beta_1$   &  $\beta$  &  $\beta$  \\ \hline
			\multirow{2}{*}{{ }}& & { } & { }&& \\
			\multirow{2}{*}{\textbf{Recovery Determinant}}& &\mbox{bad} & \mbox{good}&& \\ &  \textbf{Bayes}  & \textbf{Bayes}  &\textbf{Bayes}  & \textbf{OLS} &  \textbf{QMLE} \\ \hline 
			&&&&&\\
			\textbf{\textit{Loan characteristics}}&&&&&\\
			&&&&&\\
			(1) LOANSIZE & & \textbf{--} & \textbf{--} & &\\
			(2) LOANTYPE & \textbf{--} &  & \textbf{--} & \textbf{--} & \textbf{--}\\
			(3) LOANTYPE$\times$FIRMSIZE & & & \textbf{+} & \textbf{+} & \textbf{+} \\
			(4) ALLASSETCOLL & \textbf{+} & & \textbf{+} & \textbf{+} & \textbf{+}\\
			(5) INVENTRECIVECOLL & \textbf{+} & \textbf{+} & \textbf{+} & \textbf{+} & \textbf{+} \\				
			(6) OTHERCOLL & & \textbf{+} & \textbf{+} & \textbf{+} & \textbf{+} \\ 
			\hline &&&&&\\
			\textbf{\textit{Recovery process characteristics}}&&&&&\\
			&&&&&\\
			(7) PREPACK & & & & & \\
			(8) RESTRUCTURE & & & & & \\
			(9) OTHERDEFAULT & & & & & \\
			(10) TIMETOEMERGE & & \textbf{--} & \textbf{--} & \textbf{--} & \\
			(11) TIMETOEMERGE$^2$& & \textbf{+} & & \textbf{+} & \\
			(12) PREPACK$\times$TIMETOEMERGE & & \textbf{--} & \textbf{+} & \textbf{+} & \textbf{+} \\
			\hline &&&&&\\
			\textbf{\textit{Borrower characteristics}}&&&&&\\
			&&&&&\\
			(13) FIRMSIZE & & \textbf{--} & \textbf{+} & & \\
			(14) FIRMPPE & \textbf{+} & \textbf{+} & & & \\
			(15) FIRMCF & & & & & \\
			(16) FIRMLEV & & & & \textbf{--} & \\		
			(17) EVERDEFAULTED & \textbf{+} & \textbf{+} & \textbf{+} & \textbf{+} & \textbf{+} \\ \hline		&&&&&\\	
			\textbf{\textit{Macroeconomic \& industry conditions}}&&&&&\\
			&&&&&\\
			(18) GDP & & & & \textbf{+} & \textbf{+} \\
			(19) INDDISTRESS & & \textbf{--} & & & \\ \hline	&&&&&\\		
			\textbf{\textit{Probability of default}}&&&&&\\
			&&&&&\\
			(20) AIS & & \textbf{--} & & & \\ 
			\hline 
		\end{tabularx}
		\captionof{table}{\small{The sign of significant RR determinants under each of the Bayesian models and of those from \cite{khieu2012determinants}}.}
		\label{compareRRdets}
	\end{table*}

We note that only three loan characteristic determinants appear to be important for explaining RRs in bad times, whereas there is evidence that six are relevant during good times. We note that, like the OLS and QMLE results of \cite{khieu2012determinants}, the LOANSIZE determinant does not appear significant in the Bayesian static model. However once the credit cycle is incorporated this determinant does appear to be important.\footnote{Although significant in both `good' and `bad' states, the magnitude of the estimated marginal impact under the `good' state is relatively small.} In the case of recovery process and borrower characteristics, we find that a lesser number of variables are important in bad times and a greater number in good times. Also, and importantly, the relationships for some variables change from negative to positive. Finally, and consistent with \cite{khieu2012determinants}, the Bayesian posterior distribution for the static model shows no relation between RR and the PD measured by AIS. However, when we allow for different economic conditions, a negative relationship is indeed found between PD and RR, but only during bad times. This finding has clear implications for countercyclical capital allocations in operational risk modelling.

We now examine these variables more closely using our dynamic model. Table \ref{tab:output} reports fully detailed numerical summaries of the static and dynamic model Bayesian posterior distributions. The results for each type of RR determinant grouping are discussed in detail over the next several subsections.
	
	\begin{table*}[h!]
		\fontsize{7}{6}\selectfont
		\begin{tabularx}{\textwidth}{|m{4.3cm} >{\centering\arraybackslash}m{3.85cm} | >{\centering\arraybackslash}m{3.8cm} >{\centering\arraybackslash}m{3.8cm}|}
            \hline &&&\\
			& \textbf{Static model} & \multicolumn{2}{c|}{\textbf{Dynamic model}}\\ 
			&  $\beta$     & $\beta_0$            &  $\beta_1$ \\ \hline &&&\\
			\multirow{2}{*}{\textbf{Standardized Parameter}} &  \textbf{MPM}  &\textbf{MPM}  &\textbf{MPM} \\ 
			& \textbf{(95\% HPD)}      &\textbf{(95\% HPD)}     &\textbf{(95\% HPD)}  \\   \hline 
			&&&\\
			\textbf{\textit{Loan characteristics}}&&&\\
	&&&\\
			 \multirow{2}{*}{(1) LOANSIZE(\$M)}& $-0.026$ & $\textbf{-0.28}$   &  $\textbf{-0.093}$  \\
			& $(-0.081,0.030)$ &  $(-0.45,-0.10)$&  $(-0.18,-0.00054)$\\
			\multirow{2}{*}{(2) LOANTYPE}& $\textbf{-0.068}$ & $-0.27$&  $\textbf{-0.26}$\\
			& $(-0.13,-0.0079)$ & $(-0.43,0.11)$&  $(-0.41,-0.11)$\\
			\multirow{2}{*}{(3) LOANTYPE$\times$FIRMSIZE}& $0.025$ & $0.0098$&  $\textbf{0.35}$\\
			& $(-0.050,0.10)$ & $(-0.11,0.13)$&  $(0.0094,0.71)$\\
			\multirow{2}{*}{(4) ALLASSETCOLL}& $\textbf{0.11}$ & $0.23$ & $\textbf{0.69}$ \\
			& $(0.032,0.19)$ & $(-0.016,0.48)$&  $(0.49,0.90)$\\
			\multirow{2}{*}{(5) INVENTRECIVECOLL}& $\textbf{0.14}$ & $\textbf{1.89}$&  $\textbf{0.37}$\\				
			& $(0.069,0.22)$ & $(0.57,5.33)$&  $(0.24,0.50)$\\
			\multirow{2}{*}{(6) OTHERCOLL}& $0.057$ & $\textbf{0.25}$&  $\textbf{0.15}$\\
			& $(-0.011,0.13)$ & $(0.095,0.41)$&  $(0.028,0.26)$\\ 
			\hline &&&\\
		\textbf{\textit{Recovery process characteristics}}&&&\\
			&&&\\
	\multirow{2}{*}{(7) PREPACK}&\ $-0.0024$ & $0.14$&   $-0.14$\\
			& $(-0.084,0.080)$ & $(-0.050,0.34)$&  $(-0.34,0.057)$\\
			\multirow{2}{*}{(8) RESTRUCTURE}& $-0.016$ & $-0.19$&  $-0.0056$\\
			& $(-0.11,0.076)$ & $(-0.45,0.07)$&  $(-0.21,0.21)$\\
			\multirow{2}{*}{(9) OTHERDEFAULT}& $-0.024$ & $-0.060$&  $0.48$\\
			& $(-0.077,0.028)$ & $(-0.18,0.068)$&  $(-0.18,4.41)$\\
			\multirow{2}{*}{(10) TIMETOEMERGE}& $-0.072$ & $\textbf{-0.65}$&  $\textbf{-0.44}$\\
			& $(-0.20,0.048)$ & $(-1.20,-0.11)$&  $(-0.86,-0.0078)$\\ 
			\multirow{2}{*}{(11) TIMETOEMERGE$^2$}& $0.11$ & $\textbf{1.14}$&  $0.12$\\
			& $(-0.0093,0.24)$ & $(0.53,1.77)$&  $(-0.13,0.47)$\\ 			
			\multirow{2}{*}{(12) PREPACK$\times$TIMETOEMERGE}& $0.0064$ & $\textbf{-0.22}$&  $\textbf{0.45}$\\
			& $(-0.076,0.090)$ & $(-0.39,-0.057)$&  $(0.16,0.81)$\\ \hline &&&\\
			\textbf{\textit{Borrower characteristics}}&&&\\
			&&&\\
			\multirow{2}{*}{(13) FIRMSIZE}& $-0.022$ & $\textbf{-0.055}$&  $\textbf{0.23}$\\
			& $(-0.095,0.050)$ & $(-0.16,-0.053)$&  $(0.079,0.39)$\\
			\multirow{2}{*}{(14) FIRMPPE}& $\textbf{0.071}$ & $\textbf{0.52}$& $-0.016$ \\
			& $(0.011,0.13)$ & $(0.29,0.75)$&  $(-0.076,0.044)$\\
			\multirow{2}{*}{(15) FIRMCF}& $-0.0065$ & $0.17$&  $-0.17$ \\
			& $(-0.057,0.045)$ & $(-0.36,0.70)$&  $(-0.37,0.032)$\\
			\multirow{2}{*}{(16) FIRMLEV}& $-0.016$ & $-0.14$&  $-0.21$\\
			& $(-0.072,0.038)$ & $(-0.34,0.065)$&  $(-0.42,0.0050)$\\
			\multirow{2}{*}{(17) EVERDEFAULTED}& $\textbf{0.13}$ & $\textbf{0.53}$&  $\textbf{0.60}$\\
			& $(0.041,0.22)$ & $(0.28,0.80)$&  $(0.37,0.87)$\\ \hline &&&\\
			\textbf{\textit{Macro-eco \& industry conditions}}&&&\\
			&&&\\
			\multirow{2}{*}{(18) GDP}& $-0.0072$ & $-0.075$&  $-0.042$\\
			& $(-0.065,0.050)$ & $(-0.32,0.18)$&  $(-0.25,0.16)$\\ 
			\multirow{2}{*}{(19) INDDISTRESS}& $-0.015$ & $\textbf{-0.15}$&  $0.077$\\
			& $(-0.070,0.039)$ & $(-0.30,-0.0071)$ & $(-0.027,0.19)$\\ \hline &&&\\
			\textbf{\textit{Probability of default}}&&&\\
			&&&\\
			\multirow{2}{*}{(20) AIS}& $-0.013$ & $\textbf{-0.28}$&  $-0.018$\\
			& $(-0.070,0.043)$ &$(-0.48,-0.080)$ & $(-0.17,0.13)$\\
			\hline 
		\end{tabularx}
		\caption{Bayesian estimates of the regression coefficients based on 100,000 retained MCMC draws (with 5,000 burn-in) from each marginal posterior as indicated by the column heading. Results in the dynamic case shown in column three and four correspond to estimates conditioned on the latent credit cycle state, with $\boldsymbol{\beta_0}$ corresponding to the `bad' state, and $\boldsymbol{\beta_1}$ corresponding to the `good' state. MPM denotes the marginal posterior mean and 95\% HPD (in parentheses) denotes the 95\% higher posterior density interval. For the static model, the MPM of the squared shrinkage parameter is $\lambda^2 = 3.33$ the static case, whereas for the dynamic case, the (conditional) MPMs are $\lambda^2_0 = 2.98$ and $\lambda^2_1 = 2.71$, corresponding to the bad and good states, respectively.}
		\label{tab:output}
	\end{table*}

	\subsubsection{Loan characteristics}
	In line with \cite{dermine2006bank}, we find loan size to be negatively associated with RRs. Irrespective of being in a good or bad cycle, from a bank's perspective, the larger the loan amount, the less likely the bank will be able to recover subsequent to default. Larger loans are generally organized around a syndicate banking arrangement; hence, as more providers are involved, lower RRs are realized once they enter foreclosure. This finding is contrary to those of \cite{acharya2007does}, as banks granting larger loans are meant to have less asymmetric information and more bargaining power during the bankruptcy process. However, it seems that as loan sizes increase and default occurs during a downturn, banks are less likely to recover their outstanding debts.
	
    Loan type is not useful to explain RR levels in bad times, irrespective of whether the credit granted is a term loan or a revolver. During an upturn, however they can contribute to explaining RRs with respect to revolver loans. \cite{khieu2012determinants} find a similar significant relationship and argue that since revolvers typically have a shorter duration and are therefore reviewed more often, banks are able to reassess their clients’ credit profiles and seek further collateral if necessary.
	
	The literature emphasizes the importance of collateral with respect to higher RRs emanating from secured loans where more secured loans imply higher RRs (\cite{altman1996almost}, \cite{araten2004measuring} and \cite{van1999recovering}). Our study contributes to the literature by showing that during good times, we report similar results to those of \cite{khieu2012determinants}, i.e., a significant positive relation between the RR and total assets used as collateral. While during bad times this association is not significant, our dynamic model also shows that assets such as inventory, receivables and other more liquid assets do appear to be important for recovering a higher RR across both credit cycle states, particularly during bad times.
	
	\subsubsection{Recovery process characteristics}
	The existence of pre-arranged recovery processes for bankruptcy and out-of-court restructuring in the event of default-triggered failure is examined. Pre-packaged processes do not have a significant relationship with RRs subsequent to default in either good or bad times. Although this finding is in line with those of \cite{khieu2012determinants}, we note that the literature finds companies that pre-package appear to be more financially sound (\citet{ryan2008fair}). With respect to distressed exchanges, it transpires that firms undertaking pre-packaging are normally more solvent at the time of re-organization than are bankrupt firms (\citet{franks1994comparison}).
	
	While none of our static results support associations between any recovery process RR characteristics, the dynamic model results do indicate that TIMETOEMERGE, is negatively related to the ultimate RR, with banks less likely to recover when they engage further in bankruptcy proceedings. Owing to the magnitude of the estimated coefficient for this variable and that of its squared value, TIMETOEMERGE$^2$, it appears this relationship becomes more relevant during bad times. Furthermore, the dynamic model also finds that the constructed determinant given by PRE-PACK $\times$ TIMETOEMERGE, is also important for explaining RR outcomes irrespective of the state of credit cycle.
	\\
	
	However, the quadratic term for time to emerge, representing the nonlinearity between TIMETOEMERGE and RR, has a significant impact on recovery outcomes only in downturns. This finding is different from that observed in bond studies such as \cite{covitz2006longer}.
	
	\subsubsection{Borrower characteristics}
	The literature is not definitive on whether firm size impacts RRs. Large firms may signal higher bankruptcy costs, resulting in lower RRs. Conversely, larger firms are expected to present less information asymmetry problems to creditors, hence facilitating any restructuring process and improving recoveries from lenders. As per \cite{khieu2012determinants}, we do not find a significant relation between firm size and RRs with the static model. However, our dynamic model reveals a significant negative (positive) relation with RRs and firm size during bad (good) times. During bad times, the larger the firm, the greater the negative impact on RRs. In good times, this is reversed; larger firms are associated with greater RRs. This could be a sign of loan mispricing: recoverable assets are being over-valued prior to bankruptcy in bad times. Conversely, during good times, these asset values are more likely realizable and consistent with higher RRs.
	
	 The level of a firm's tangible assets, namely property, plant and equipment (FIRMPPE), is thought to be positively related to the RR (\cite{acharya2007does}), that is, banks are more likely to recover outstanding loans when firms report tangible assets on their balance sheet. We find likewise, but only relating to bad times. (We note that \cite{acharya2007does} include bonds that are generally unsecured, in their sample.) We also find that firm cash flow and leverage are not significantly related to RRs. Unlike  \cite{khieu2012determinants}, as our dataset excludes bonds and focuses only on loans (which are likely to be secured by tangible assets), this contrast is not surprising. Finally, consistent with the literature, we find that prior defaults as indicated by the variable EVERDEFAULTED are significant and positively related to RRs in both good and bad times.
	
	\subsubsection{Macroeconomic \& Industry conditions}
	We find no significant relation between GDP and RRs, however this is due to the fact that we control for the underlying economic conditions, which coincides with the credit cycle (see Figure \ref{creditcycle} and the discussion in Section \ref{5.2}). We do, however, obtain a significant negative relation between industry distress (measured by stock returns, via the INDDISTRESS variable) and RRs in bad times.
	
	\subsubsection{Probability of default}
	The AIS is used to measure the PD, following \cite{khieu2012determinants}. The literature suggests a negative relationship between the PD and the RR. Both \cite{hu2002dependence} and \cite{altman2005link} report a negative association, although the former uses bond default data. \cite{khieu2012determinants} report no relationship between the PD and the RR. We find that PD is significantly negatively related to RR, but only in bad times, arguably consistent with banks being less likely to recover under an increased PD during bad times. This finding is partially consistent with \cite{altman2005link}, although the study does not distinguish between good and bad times.
	
	\section{Conclusion} \label{Section 6}
	Using US bank default loan data from Moody's Ultimate Recovery Database and covering the pre- and post-GFC period, this paper develops a dynamic predictive model for bank loan RRs, accommodating the distinctive features of the empirical RR distribution and incorporating a large number of possible determinants. Furthermore, some of the factors that are analyzed and reported in the literature have been overlooked as insignificant, due to the static model approach, which does not control for the different states of the economy. Our temporal conditioning in a hierarchical framework allows us to discriminate between good and bad states of the credit cycle. Thus, this paper contributes to the literature in different ways. The methodological approach used is Bayesian in nature and therefore is able to handle the hierarchical specification that is built to explain the complex relationship between PD, determinants and the empirical distribution of RRs. It is the first paper to incorporate time-series variation into the probabilistic modelling of bank loan RRs, proposing a Bayesian hierarchical framework that enables inference of a latent credit cycle. We also introduce the use of a LASSO prior to encourage the most relevant RR determinants to be found, despite potentially confounding evidence of correlation between observed RR determinants.
	
	We find that some loan characteristics such as those using specific types of collateral hold different explanatory power in good times and bad. We find that certain recovery process variables, such as the length of time between default and resolution for loans with pre-packaged recoveries, differ in their importance in relation to RRs, depending on the state of the credit cycle, in this case being negatively related to RR in bad times while being positively related in good times. Only a few borrower characteristics and industry conditions appear to be relevant across the cycle. The defaulting firm’s size and asset tangibility can imply different relationships with RRs depending on conditions. Finally, by allowing for variation in the level of PD, on top of the latent dynamic cycle states, we find a negative relationship with RR but one that is only significant during credit downturns.
	
	Our results illustrate the importance of utilizing dynamic models that allow for time-varying conditions, as there is significant variation in the explanatory power of the variables analyzed depending on these conditions, yielding new insights previously unavailable from the established literature. Taking the case of PD, no relation between RR and PD is reported, yet under our dynamic model we find it is significantly and negatively related to RR during bad times. This variation in significance in variables across good and bad times occurs in several of our variables, supporting the need for a dynamic approach.
	
	Our results also yield significant implications for the banking sector, notably providing empirical support for the latest addition to the Basel framework concerning the importance of activating countercyclical capital buffers during economic downturns. Applying such a buffer would not only enable banks to absorb increased losses but would also assist in achieving the broad macro-prudential goals of protecting the banking sector in periods of excess aggregate credit growth, and from the build-up of system-wide risk.
	
	The notion that RR is driven by a systemic risk component that becomes more pronounced during bad times is evident from the results reported in our dynamic model. Loan size and type are also critical features, especially during bad times. Such features need to be priced within the cost of financing, as some banks are less likely to recover when the economy is entering a downturn. These important differential impacts, during bad and good times, suggests that RRs have a large element of systemic risk that needs to be factored in during the pricing of loan finance contracts. As RRs are an integral part of credit risk, this aspect should attract an additional risk premium allowing for a differentiated credit risk exposure.
	
	Under the new regime, banks are required to provide more timely and forward-looking information. It is no longer necessary for a credit event to have occurred before a credit loss is recognized. The paradigm shift is in being cognizant of the credit cycle and to update the bank’s loan loss provision in line with their recovery rate expectations.
	
	In summary, we find several variables are important for explaining RRs, depending on the state of the credit cycle. This has major implications for the countercyclicality of regulatory capital and operational risk management. The potential risk of not addressing such factors will result in either underestimating the relevant credit risk, or overestimating it. Both of these eventualities could potentially result in negative consequences, such as more expensive loans. This in turn would result in desirable customers leaving to access financing at cheaper rates from alternative institutions more effective in correctly pricing loans through the procyclical process.
	
	\begin{appendices}

	\section{The definitions of the recovery rate determinants}
	Table \ref{tab:rrdets} details each of the twenty RR determinants considered in this paper, with the name of the determinant given in the first columns and the corresponding definition given in the second column. The determinants are clustered according broad type (loan characteristics, recovery process characteristics, borrower characteristics, macroeconomic and industry condition determinants and the probability of default) and each given a unique determinant number (in parentheses preceeding the name) that is used throughout the paper.
	
		\begin{table*}[h!]
			\fontsize{8}{7}\selectfont
			\begin{tabularx}{\textwidth}{|p{5cm}|p{12cm}|}
				\hhline{--} & \\
				\textbf{Name}     & \textbf{Definition} \\ \hhline{--}
				\vspace{0.01 cm} \textbf{\textit{Loan characteristics}}&\\
				\vspace{0.01 cm} (1) LOANSIZE(\$M)     & \vspace{0.01 cm}  The dollar amount (in millions of dollars) of the facility at the time of issuance.   \\
				&	\\
				(2) LOANTYPE                   & A dummy variable equal to one if the loan is a term loan (fixed tenure and not recallable on demand), and equal to zero if it is a revolver (short-term revolving and recallable on demand).  \\
				&	\\
				(3) LOANTYPE $\times$ FIRMSIZE   &  The product of LOANTYPE and FIRMSIZE \\
				&	\\
				(4) ALLASSETCOLL               &  A dummy variable equal to one if the loan is secured by all firm assets, and zero otherwise.  \\
				&	\\
				(5) INVENTRECIVECOLL           &  A dummy variable equal to one if the loan is secured by inventory, accounts receivable, or both, and zero otherwise. \\
				&	\\
				(6) OTHERCOLL                  &  A dummy variable equal to one if the loan is secured differently from the other types, and zero otherwise.\\
				\hhline{--}
				\vspace{0.01 cm} 	\textbf{\textit{Recovery process characteristics}}&\\
				\vspace{0.01 cm} (7) PREPACK     & 	\vspace{0.01 cm}  A dummy variable equal to one if the bankruptcy is through a pre-packaged bankruptcy, and zero otherwise.\\
				&	\\
				(8) RESTRUCTURE              &  A dummy variable equal to one if default is resolved by out-of-court restructuring, including distressed exchange offers, and zero otherwise. \\
				&	\\
				(9) OTHERDEFAULT             &  A dummy variable, equal to one if default is resolved by other methods than an out-of-court restructuring, pre-packaged  formal bankruptcy, and 0 otherwise. \\
				&	\\
				(10) TIMETOEMERGE             &  The length of time (in months) between bankruptcy or restructuring and emergence, often known as resolution time. \\
				&	\\
				(11) TIMETOEMERGE$^2$         &  TIMETOEMERGE squared. \\
				&	\\
				(12) PREPACK$\times$ TIMETOEMERGE   & The product of PREPACK and TIMETOEMERGE. \\
				\hhline{--} 
				\vspace{0.01 cm} \textbf{\textit{Borrower characteristics}}&\\
				\vspace{0.01 cm} (13) FIRMSIZE            &  				\vspace{0.01 cm}  The market value of firm-level assets one year before default. The market value is calculated as the book value of long-term and short-term debt plus the number of common shares outstanding. \\
				&	\\
				(14) FIRMPPE             &  Firm asset tangibility, measured as net property, plant, and equipment over total book assets one year before default. \\
				&	\\
				(15) FIRMCF              &  Firm cash flows, measured EBITDA (earning before interest, tax and depreciation and amortization) over total book assets one year before default. \\
				&	\\
				(16) FIRMLEV             &  Firm leverage, measured as total long-term debt plus debt in current liabilities over total book assets one year before default.  \\
				&	\\
				(17) EVERDEFAULTED       &  A dummy variable equal to one if the firm has defaulted before, and zero otherwise. \\
				\hhline{--}
				\vspace{0.01 cm} 	\textbf{\textit{Macro-eco \& industry conditions}}&\\
				\vspace{0.01 cm} (18) GDP            &   \vspace{0.01 cm} The annual GDP growth rate measured 1 year before default. \\
				&	\\
				(19) INDDISTRESS    &  A dummy variable equal to one if the industry median stock returns in the year default is less than -30\%, and zero otherwise. The stock returns are calculated without the defaulting firms and the industry is defined according to the three-digit SIC codes. \\
				\hhline{--}
				\vspace{0.01 cm} 	\textbf{\textit{Probability of default}}&\\
				\vspace{0.01 cm} (20) AIS    & 		\vspace{0.01 cm}  The credit spread (in percent) at the time of loan origination over LIBOR of the drawn loan that defaulted. \\
				\hhline{--}
			\end{tabularx}
			\captionof{table}{\small Definitions of the RR determinants.}
			\label{tab:rrdets}
		\end{table*}

		\section{Implementation details for Bayesian analysis} \label{Appendix A}
		The model detailed in Section \ref{Section 4} provides a characterization of the distribution of the observed RRs via a predictive regression of the RR mixture component on a large collection of RR determinants, with the regression coefficients in turn dependent upon the current state of an underlying credit cycle state variable. In the static regression setting, calculation of the posterior distribution via MCMC follows the approach of \cite{altman2014ultimate}, who in turn rely upon the methodology details provided in \cite{albert1993bayesian}. Our implementation here is similar, including in the dynamic case where we include the additional hierarchical layer containing the Markov switching variables, except for the use of the alternative LASSO prior specification on the latent RR regression parameter coefficient vector(s).
		
		\subsection{Likelihood function}
		\label{posterior}
		The relevant likelihood for Bayesian analysis is the joint probability density function (pdf) of the complete set of measurements, denoted by $\mathbf{y}=(y_1,y_2,\ldots,y_n)$, together with the latent Markov switching state variables, $\mathbf{S} = (S_{1},S_{2},\ldots,S_{T}),$ all conditional upon the collection of parameters,  $(\boldsymbol{\mu}, \boldsymbol{\sigma^2}, \boldsymbol{\beta_0}, \boldsymbol{\beta_1})$. Unfortunately, even if the sequence of latent credit state variables, $\mathbf{S}$, were known, calculation of the likelihood function is not available in closed form, and consequently the Bayesian posterior is also not available. However, owing to the relationship between the Gaussian mixture model for each $\mathbf{y_i}$, the cut-points $\mathbf{c}$ and the latent predictive regression in (\ref{Eq_CF2}), we can express the joint pdf of $\mathbf{y}$ and $\mathbf{z}$ conditional on $\mathbf{S}$, given by the product of
		\begin{align}
		p(\mathbf{y},\mathbf{z}, \mathbf{z^{\ast}}|\mathbf{S},\boldsymbol{\psi}, \mathbf{x}) & \propto \prod_{i=1}^n \prod_{j=1}^{J} \nonumber \frac{1}{\sigma_j}\phi(\frac{y_i - \mu_j}{\sigma_j})\\
		& \times \mathbb{I}(c_{j-1}<z_i\leq c_j)	\times \mathbb{I}(z^{\ast}_i =j) \nonumber\\ 
		& \times \phi(z_i - \mathbf{x}_i'((1-S_{t_i})\boldsymbol{\beta_0}+S_{t_i}\boldsymbol{\beta_1})),  	\label{Equation 5}
		\end{align}
		where $\phi(\cdots)$ denotes the pdf of the standard normal distribution, $\mathbb{I}(\cdot)$ is the indicator function so that $\mathbb{I}(A)=1$ if event $A$ is true and is equal to zero otherwise, $\boldsymbol{\psi} = (\boldsymbol{\mu}, \boldsymbol{\sigma^2},  \boldsymbol{\beta_0}, \boldsymbol{\beta_1}, \mathbf{c})$, and the joint pdf of the Markov switching states $\mathbf{S}$, given by
		\begin{equation}
		p(\mathbf{S}|\boldsymbol{\psi}) = p(S_1|p,q)\prod_{t=2}^T p(S_t|S_{1:t-1},p,q),
		\label{Equation 6}
		\end{equation}
		where $p(S_t|S_{1:t-1},p,q)$ is given in (\ref{MSprob}), and $p(S_1|p,q)$ arising from the long-run marginal probability given by 
		\begin{equation}
		S_{1}|p,q \sim Bernoulli((1-p)/(2-p-q)).
		\end{equation}
		The likelihood function is then the product of  (\ref{Equation 5}) and (\ref{Equation 6}). Note that the factors in these equations are expressed  conditionally given the parameters $(\boldsymbol{\psi},p,q)$ and also given the regression covariates $\mathbf{x}=\left[\mathbf{x}_1,\mathbf{x}_2,\ldots,\mathbf{x}_K \right]$, where $\mathbf{x}_k=(x_{k,1},x_{k,2},\ldots,$ $ x_{k,n})'$.
		
		\subsection{Priors}\label{sec:priors}
		
		To complete a fully Bayesian analysis, we must put a (joint) prior distribution over the unknown parameters. We take these priors to be relatively diffuse, so that the data will dominate the analysis. Specifically, the prior mean $\mu_j$ and variance $\sigma^2_j$ for the $j^{th}$ mixture component of the RR distribution are taken as independent normal ($N$) and inverse gamma ($IG$) distributions, with
		\begin{itemize}
			\item  $\mathbf{\mu}_j \overset{ind}{\sim} N(\bar{\mu}_j \ , \ \bar{V}_{\mu,j})$, where $\bar{\mu}_j =0$ and $\bar{V}_{\mu,j} = 100$ for $j=1,\cdots,J$, and
			\item  $\sigma^2_j  \overset{ind}{\sim} IG(\bar{a}_j \ , \ \bar{b}_j)$, where $\bar{a}_j=3$ and $\bar{b}_j=1$ denote the scale and shape parameters, respectively, for $j = 1,\cdots,J$.
		\end{itemize}
		To avoid the well known label switching problems in the finite mixture model, we impose the same identification restrictions, $\mu_1 < \cdots < \mu_J$, as in \citet{koop2007bayesian}.\footnote{For a detailed discussion of alternative solutions to the label switching problem, see \citet{fruhwirth2006finite}.} The joint prior distribution for the cut-point vector $\mathbf{c}$ is completely diffuse, while the prior distributions for $\boldsymbol{\beta_0}$ and $\boldsymbol{\beta_1}$, corresponding to the Bayesian LASSO for the coefficients of the RR determinants under the `bad' and `good' credit cycle states, respectively, are specified hierarchically using independent scale mixture of normals for each. These are given by 
		\begin{itemize}
			\item $\boldsymbol{\beta_0} \mid \sigma^2_{\varepsilon}, \boldsymbol{\tau_0} \sim N(\mathbf{0}_K,\sigma_{\varepsilon}^2 \mathbf{I}_KD_{\boldsymbol{\tau_0}})$,  with \\ $D_{\boldsymbol{\tau_0}} = diag(\tau_{0,1},\tau_{0,2},\ldots,\tau_{0,K})$, and 
			\item $\boldsymbol{\beta_1} \mid \sigma^2_{\varepsilon}, \boldsymbol{\tau_1} \sim N(\mathbf{0}_K,\sigma_{\varepsilon}^2 \mathbf{I}_KD_{\boldsymbol{\tau_1}})$,  with \\ $D_{\boldsymbol{\tau_1}} = diag(\tau_{1,1},\tau_{1,2},\ldots,\tau_{1,K})$, 
		\end{itemize}
		with $I_K$ denoting the $K-$ dimensional identity matrix and the mixing variables (also known as local shrinkage parameters) given by $\boldsymbol{\tau_0} =(\tau_{0,1},\tau_{0,2},\ldots,$ $\tau_{0,K})$ and $\boldsymbol{\tau_1} =(\tau_{1,1},\tau_{1,2},\ldots,\tau_{1,K})$ and with the variance $\sigma_{\varepsilon}^2 = 1$ held fixed as used in the latent ordered probit regression. In addition, following \cite{park2008bayesian}, we use the following independent (hyper) priors
		\begin{itemize}
			\item $\tau_{0,1}, \tau_{0,2}, \ldots, \tau_{0,K} \mid \lambda_0^2 \overset{ind}{\sim} Exp(\lambda_0^2/2), $ and
			\item $\tau_{1,1}, \tau_{1,2}, \ldots, \tau_{1,K} \mid \lambda_1\overset{ind}{\sim} Exp(\lambda_1^2/2), $ 
		\end{itemize}
		where $Exp(s)$ denotes the exponential distribution with mean value $1/s$. Then, the (hyper) prior for the two global LASSO parameters $\lambda_0^2$ and $\lambda_1^2$ is given by independent distributions
		\begin{itemize}
			\item $\lambda^2_0 \sim \mbox{Gamma}(\bar{r},\bar{\delta})$, and 
			\item $\lambda^2_1 \sim \mbox{Gamma}(\bar{r},\bar{\delta})$, 
		\end{itemize}
		where $\bar{r} = 3$ and $\bar{\delta} = 1$. Finally, we have priors for the Markov switching probabilities, corresponding to the parameters in (\ref{MSprob}), given by
		\begin{itemize}
			\item $p \sim \mbox{Beta}(\bar{u}_{0,0},\bar{u}_{0,1})$ and $q \sim \mbox{Beta}(\bar{u}_{1,0},\bar{u}_{1,1})$, 
		\end{itemize}
		with $\bar{u}_{0,0}$, $\bar{u}_{0,1}$, $\bar{u}_{1,0}$ and $\bar{u}_{1,1}$ all set equal to 0.5, as per the algorithm of \cite{kim2001bayesian}. Collectively, the joint prior distribution is specified over the entire collection of unknown parameters $ \boldsymbol{\theta} = (\boldsymbol{\psi},p,q, \boldsymbol{\tau_0},\boldsymbol{\tau_1}, \lambda_0,\lambda_1^2) $ is given by
		\begin{align*}
		p(\boldsymbol{\theta}) & = p(\boldsymbol{\mu}) \ p(\boldsymbol{\sigma^2}) \ p(\mathbf{c}) \\ 
		& \times p(\boldsymbol{\beta_0} \mid \boldsymbol{\tau_0})
		\ p(\boldsymbol{\beta_1} \mid \boldsymbol{\tau_1})  \\
		& \times p(\boldsymbol{\tau_0} \mid \lambda^2_0) \  p(\boldsymbol{\tau_1} \mid \lambda^2_1) \  p(p) \ p(q).
		\end{align*}

		\subsection{MCMC Algorithm}
		The marginal posterior distribution is obtained using a basic Gibbs sampling approach, where the parameters and latent variables are each drawn recursively from the relevant (full) conditional posteriors. Given the prior distribution, the $g^{th}$ iteration of the Gibbs sampler proceeds as follows: 
		\begin{itemize}
			\item Step 1: draw the mixture indicators and the predictive scores\\ $\mathbf{z^{\ast(g)}}, \mathbf{z^{(g)}} | \mathbf{y}, \boldsymbol{\mu}^{(g-1)},\boldsymbol{\sigma}^{2(g-1)}, \mathbf{c}^{(g-1)},\boldsymbol{\beta_0}^{(g-1)},\boldsymbol{\beta_1}^{(g-1)},\\ \hspace*{\fill} \mathbf{c}^{(g-1)}, \mathbf{S}^{(g-1)} $
			\item Step 2: draw the regression parameters\\ $\boldsymbol{\beta}_0^{(g)}|\mathbf{y},\mathbf{z}^{(g)},\mathbf{S}^{(g-1)},\boldsymbol{\tau}^{(g-1)}$ and \\
			$\boldsymbol{\beta}_1^{(g)}|\mathbf{y},\mathbf{z}^{(g)},\mathbf{S}^{(g-1)},\boldsymbol{\tau}^{(g-1)}$ 
			\item Step 3: draw the shrinkage parameters (via augmentation) \\ $\lambda_0^{2(g-1)},\tau_0^{(g)} \mid \boldsymbol{\beta}_0^{(g)}$ and \\
			$\lambda_1^{2(g-1)},\tau_1^{(g)} \mid \boldsymbol{\beta}_1^{(g)}$ 
			\item Step 4: draw each of the $J$ cut-points\\ $c^{(g)}_j|\mathbf{z}^{(g)},\bm{c}_{/j}$ 
			\item Step 5: draw the latent Markov states\\ $\mathbf{S}^{(g)}|y,\mathbf{z}^{(g)},\boldsymbol{\beta_0}^{(g)},\boldsymbol{\beta_1}^{(g)},p^{(g-1)},q^{(g-1)}$
			\item Step 6: draw the Markov transition probabilities\\ $p^{(g)},q^{(g)}|\mathbf{S}^{(g)}$ 
			\item Step 7: draw the vector of mean parameters for the Gaussian mixture distribution\\ $\boldsymbol{\mu}^{(g)}|y,\mathbf{z}^{(g)},\bm{c}^{(g)},\boldsymbol{\sigma}^{2(g-1)}$ and $\boldsymbol{\sigma}^{2(g)} \mid y, \mathbf{z}^{(g)},\bm{c}^{(g)},\boldsymbol{\mu}^{(g)}$.
			\item Step 8: draw the vector of variance parameters for the Gaussian mixture distribution\\ 
			$\boldsymbol{\sigma}^{2(g)} \mid y, \mathbf{z}^{(g)},\bm{c}^{(g)},\boldsymbol{\mu}^{(g)}$.
		\end{itemize}
		
		We note that in Step 3 each shrinkage parameter, either $\lambda_0^2$ and $\lambda_1^2$ (or $\lambda^2$ in the static case), is generated by first sampling an augmentation vector, $\tau_0^{(g)}$ and $\tau_0^{(g)}$, respectively, from the corresponding distribution that conditions on the relevant previous draw of the shrinkage parameter. This approach follows as per \cite{park2008bayesian}. The new draws of the shrinkage parameters are then sampled from the full conditional distributions that utilize the augmentation vectors, i.e. $\lambda_0^{2(g)} \sim p(\lambda_0^{2} \mid \tau_0^{(g)}, \boldsymbol{\beta}_0^{(g)})$ and $\lambda_1^{2(g)} \sim p(\lambda_1^{2} \mid \tau_1^{(g)}, \boldsymbol{\beta}_1^{(g)})$, respectively. The values of the $\tau_0^{(g)}$ and $\tau_1^{(g)}$ are not required for any additional part of the MCMC algorithm and may be discarded at the end of each iteration. We also note for Step 4 that $\bm{c}_{/j}=\{c_0,c_1,\ldots,c_{j-1},c_{j+1},\ldots,c_J\}$, denoting the vector $\mathbf{c}$ but with the $j^{th}$ element excluded. Since the priors are conditionally conjugate for all unknowns, the relevant conditional posterior distributions are all derived analytically, ensuring a fast algorithm for sampling from the full joint posterior distribution.
		
		\subsection*{\bfseries Mixture indicator vector $\bm{z^\ast}$ and the latent predictive score vector $\bm{z}$}
		The mixture indicator variable components $z^{\ast}_1,\cdots,z^{\ast}_N$ are conditionally independent and \textbf{hence are} be sampled independently from multinomial distributions with probabilities
		\begin{align*}
		p(z_i^{\ast} &|\mathbf{y},\boldsymbol{\mu}',\boldsymbol{\sigma}^{2'}, \bm{c},\boldsymbol{\beta_0},\boldsymbol{\beta_1}) \\ &\propto p(z_i^{\ast}|w_i)p(w_i|\mathbf{y},\boldsymbol{\mu},\boldsymbol{\sigma}^{2'}, \bm{c},\boldsymbol{\beta_0},\boldsymbol{\beta_1}),
		\end{align*}
		with diffuse priors on $\bm{w}$, we get
		\begin{align*}
		w&_{i,j}= & \\ 
		& \frac{\left[\Phi(\textbf{x}'_i\beta_{S_{t_i}}-c_{j-1})-\Phi(\textbf{x}'_i\beta_{S_{t_i}}-c_{j})\right] \ \phi(y_i;\mu_j,\sigma^2_j)}{\sum_{j=1}^{J}\left[\Phi(\textbf{x}'_i\beta_{S_{t_i}}-c_{j-1})-\Phi(\textbf{x}'_i\beta_{S_{t_i}}-c_{j})\right] \ \phi(y_i;\mu_j,\sigma^2_j)},
		\end{align*}
		for $i = 1,\cdots,n$ and $j = 1,\cdots,J$, where $\Phi(\cdot)$ denotes the cdf of a standard normal random variable. 
		
		Conditional on the sampled mixture indicator variable $z_i^{\ast}$ (as well as on the relevant latent credit state coefficient $\boldsymbol{\beta_{S_{t_i}}}$ corresponding to the latent credit state at the time of default for $RR_i$, the data $\mathbf{y}$, other parameters), the latent data for individual $i$, given by $z_i$, is generated from a truncated normal distribution 
		\begin{equation*}
		p(z_i|\textbf{c},\beta_0,\beta_1,z^\ast_i) \sim \mathcal{TN}_{(c_{z^\ast_i-1},c_{z^\ast_i})} (\textbf{x}'_i\beta_{S_{t_i}},1),
		\end{equation*}
		where $c_{z^\ast_{i-1}}$ and $c_{z^\ast_i}$ are the lower bound and the upper bound parameters. 
		
		\subsection*{\bfseries State-dependent regression coefficients $\boldsymbol{\beta_0}$ and $\boldsymbol{\beta_1}$}
		The latent data $\mathbf{z}$ and recovery determinants $\bm{x}$ are divided into $\mathbf{Z}_{0}$, $\mathbf{Z}_{1}$ and $\bm{X}_0 = [\bm{x}_{1,0},\bm{x}_{2,0},\cdots,\bm{x}_{k,0}]$, $\bm{X}_1 = [\bm{x}_{1,1},\bm{x}_{2,1},\cdots,\bm{x}_{k,1}]$ respectively, according the latent \newline Markov state $S_{t_i}$. Given the data, $\mathbf{y}$ and $\mathbf{z}$, Markov states, $S_{t_i}$, and the local shrinkage parameter, $\tau_0^2 = (\tau_{1,0},\cdots,\tau_{k,0})$ and $\tau_1^2 = (\tau_{1,1},\cdots,\tau_{k,1})$, the conditional posterior distribution for $\beta_{S_{t_i}} = (\beta_0,\beta_1)'$ is given by
		\begin{equation*}
		p(\beta_{S_{t_i}}|\mathbf{y},\mathbf{z},\mathbf{S},\tau) \sim \mathcal{N}(D_{S_{t_i}}d_{S_{t_i}}, D_{S_{t_i}}),
		\end{equation*}
		where 
		\begin{align*}
		D_{S_{t_i}} = (\textbf{x}_{S_{t_i}}'\textbf{x}_{S_{t_i}} + \mbox{diag}(\tau^2)^{-1})^{-1} \ \mbox{and} \ d_{S_{t_i}} = \textbf{x}_{S_{t_i}}'y_{S_{t_i}}.
		\end{align*}
		
		\subsection*{\bfseries Shrinkage parameters $\tau^2$ and $\lambda^2$}
		For each credit state, the local shrinkage parameters $\tau^2_1,\cdots,\tau^2_k$ are conditionally independent, with
		\begin{equation*}
		p(1/\tau_j^2|\beta_j,\lambda^2) \sim InvGaussian(\bar{\bar{\mu}}_j, \lambda^2),
		\end{equation*}
		where $InvGaussian$ denotes an Inverse Gaussian distribution and
		\begin{equation*}
		\bar{\bar{\mu}}_j = \sqrt{\frac{\lambda^2}{\beta_j^2}},
		\end{equation*}
		for $j=1,\cdots,k$. With a conjugate prior, the full conditional distribution of $\lambda^2$ is given by
		\begin{equation*}
		p(\lambda^2|\tau^2) \sim \Gamma(\bar{\bar{r}},\bar{\bar{\delta}}),
		\end{equation*}
		where $\Gamma$ denotes a gamma distribution with shape parameter $K+\bar{r}$ and rate parameter $\sum_{j=1}^{K}\tau_j^2/2+\bar{\delta}$.
		
		\subsection*{\bfseries Cut-points $\bm{c}$}
		We follow \cite{albert1993bayesian} and use diffuse priors for all cut-points $c_2,\cdots,c_{J-1}$. For identification purpose, we set $c_0 =-\infty$, $c_1 =0$ and $c_J = \infty$ as it is common in any other studies using an ordered probit model. The joint conditional posterior for the cut-points $j = 2,\cdots,J-1$ (recall that $c_0 = -\infty, c_1 = 0,$ and $c_J=\infty$) is given by, 
		\begin{equation*}
		p(c_j|c_{j} ,\mathbf{z}, \mathbf{z}^\ast) \sim \mathcal{U}(l_j,u_j),
		\end{equation*}
		where $\mathcal{U}(l_j,u_j)$ denotes a uniform distribution with
		\begin{align*}
		l_j &= \max{\{c_{j-1}, \max\{z_i:z_i^\ast=j\}\}},\\
		u_j &= \min{\{c_{j+1}, \min\{z_i:z_i^\ast=j+1\}\}}.
		\end{align*}
		
		\subsection*{\bfseries The latent Markov states S}
		We use the efficient block sampling algorithm of \cite{carter1994gibbs} and \cite{fruhwirth1994data} to generate $\mathbf{S}$, which is known as \textit{forward filtering, backward sampling} (FFBS). \cite{hamilton1989new} provides the following filtering algorithm to calculate the filtered probabilities for $\mathbf{S}$. Let $z_t$ be vector contains all $z_i$ observed in year $t$, the Hamilton filter consists of, for $t=1,\cdots,T$,
		\begin{itemize}
			\item predict
			\begin{align*}
			\begin{bmatrix} \Pr(S_{t}=0|z_{t}) \\ \Pr(S_{t}=1|z_{t})	\end{bmatrix}
			=\begin{bmatrix} p & 1-q \\ 1-p & q	\end{bmatrix}\begin{bmatrix} \Pr(S_{t}=0|z_{t-1}) \\ \Pr(S_{t}=1|z_{t-1})	\end{bmatrix},
			\end{align*}
			\item update
			\begin{align*}
			\Pr(S_{t}=0|z_t) &\propto p(z_t|S_{t} =0)\Pr(S_{t}=0|z_{t}) \mbox{ and }\\
			\Pr(S_{t}=1|z_t) &\propto p(z_t|S_{t} =1)\Pr(S_{t}=1|z_{t}),
			\end{align*}
		\end{itemize}
		where
		\begin{align*}
		p(z_t|S_{t} =0) &= \prod_{i=1}^{n_t}p(z_{i_t}|S_{t} =0)  \mbox{ and }\\
		p(z_t|S_{t} =1) &= \prod_{i=1}^{n_t}p(z_{i_t}|S_{t} =1).
		\end{align*}
		The latent Markov states are then simulated sequentially, for $t=T, T-1, ..., 1$. Given $S_{t+1}$, the parameter in the Bernoulli distribution for each $t$ is calculated by 
		\begin{equation*}
		\Pr(S_{t}=1|z_{1:T})/(\Pr(S_{t}=0|z_{1:T})+\Pr(S_{t}=1|z_{1:T})),
		\end{equation*}
		where
		\begin{align*}
		\Pr(S_{t}=0|z_{1:T}) &\propto \Pr(S_{t}=0|z_{t})\Pr(S_{t}=0|S_{t+1}) \mbox{ and }\\
		\Pr(S_{t}=1|z_{1:T}) &\propto \Pr(S_{t}=1|z_{t})\Pr(S_{t}=1|S_{t+1}).
		\end{align*}

		\subsection*{\bfseries The Markov transition probabilities $p$ and $q$}
		Conditional on $\mathbf{S}$, the transition probabilities, $p$ and $q$ are independent of the data. Since we have assigned beta prior distributions to the transition probabilities, the conditional posterior distributions are given by
		\begin{align*}
		p(p,q|\mathbf{S}) \propto p(p,q) p(\mathbf{S} \mid p,q),
		\end{align*}
		where $p(\mathbf{S}|p,q)$ describes the joint probabilities associated with the latent Markov-switching states. Prior independence (of $p$ and $q$) implies posterior independence here, and hence $p$ and $q$ may be jointly sampled according to
		\begin{align*}
		p|\mathbf{S} &\sim \mathcal{B}(\bar{u}_{0,0}+n_{0,0}, \bar{u}_{0,1}+n_{0,1}) \ \mbox{and}\\
		q|\mathbf{S} &\sim \mathcal{B}(\bar{u}_{1,1}+n_{1,1}, \bar{u}_{1,0}+n_{1,0}),
		\end{align*}
		where $\mathcal{B}(a,b)$ denotes the Beta distribution on $(0,1)$, having mean $\frac{a}{a+b}$ and variance $\frac{ab}{(a+b)^2(a+b+1)}$, here with
		\begin{align*}
		n_{1,0} = \sum_{t=1}^{T}\sum_{i=1}^{n_t}S_{t_i}|S_{t-1}=0,\\
		n_{1,1} = \sum_{t=1}^{T}\sum_{i=1}^{n_t}S_{t_i}|S_{t-1}=1,
		\end{align*}
		and $n_{0,0} = n_{S_{t}=0} - n_{1,0}$ and $n_{0,1} = n_{S_{t}=1} - n_{1,1}$.
		
		\subsection*{\bfseries The Gaussian mixture means $\boldsymbol{\mu}$ }
		Given the independent conjugate priors, the individual $\mu_j$, for $j=1,\cdots,J$, may be sampled independently from 
		\begin{equation*}
		\mu_j|\mathbf{y},\mathbf{z},\sigma_j^2  \sim  \mathcal{TN}_{(\mu_{j-1},\mu_{j+1})} (D_{\mu_j}d_{\mu_j},D_{\mu_j}),
		\end{equation*}
		where 
		\begin{equation*}
		D_{\mu_j}=\left(\sum_{i=1}^{n}\mathbb{I}(z^\ast_i=j)/\sigma^2_j+\bar{V}_{\mu_j}^{-1}\right)^{-1},
		\end{equation*}
		and
		\begin{equation*}
		d_{\mu_j}=\sum_{i=1}^{n}\mathbb{I}(z^\ast_i=j)y_i/\sigma^2_j+\bar{V}_{\mu_j}\bar{\mu}_j.
		\end{equation*}
		
		\subsection*{\bfseries The Gaussian mixture variances $\boldsymbol{\sigma^2$}}
		The individual variance parameters  $\sigma^2_j$ for mixture components $j=1,\cdots,J$, are sampled independently conditional on $\mu_j$ and $\mathbf{z^{\ast}}$ given by
		\begin{align*}
		p(\sigma_j^2|\mathbf{y},\mathbf{z^{\ast}},\mu_j) \sim IG(\bar{\bar{a}}_j,\bar{\bar{b}}_j),
		\end{align*}
		with
		\begin{align*}
		\bar{\bar{a}}_j = \frac{\sum_{i=1}^{n}\mathbb{I}(z^{\ast}_i=j)}{2}+\bar{a}_j,
		\end{align*}
		and
		\begin{align*}
		\bar{\bar{b}}_j = \bar{b}_j^{-1}+\frac{1}{2}\sum_{i=1}^{n}\mathbb{I}(z^\ast_i=j)(y_i-\mu_j)^2.
		\end{align*}
		
	\end{appendices}


\bibliography{RR}

\end{document}